\documentclass[12pt]{article}
\usepackage{latexsym}
\usepackage{graphics}
\usepackage{epsfig}

\newcommand{\bmat}{\left(\begin{array}}
\newcommand{\emat}{\end{array}\right)}
\newcommand{\beq}{\begin{equation}}
\newcommand{\eeq}{\end{equation}}

\newcommand{\drawsquare}[2]{\hbox{%
\rule{#2pt}{#1pt}\hskip-#2pt
\rule{#1pt}{#2pt}\hskip-#1pt
\rule[#1pt]{#1pt}{#2pt}}\rule[#1pt]{#2pt}{#2pt}\hskip-#2pt
\rule{#2pt}{#1pt}}

\newcommand{\Ysymm}{\raisebox{-.5pt}{\drawsquare{6.5}{0.4}}\hskip-0.4pt%
        \raisebox{-.5pt}{\drawsquare{6.5}{0.4}}}
\newcommand{\Yasymm}{\raisebox{-3.5pt}{\drawsquare{6.5}{0.4}}\hskip-6.9pt%
        \raisebox{3pt}{\drawsquare{6.5}{0.4}}}

\def\yzero{\smash{\hbox{$y\kern-4pt\raise1pt\hbox{${}^\circ$}$}}}

\def\-{\hphantom{-}}
\def\ov{\overline}
\def\s2{\frac{1}{\sqrt2}}

\def\beq{\begin{equation}}
\def\eeq{\end{equation}}
\def\beqa{\begin{eqnarray}}
\def\eeqa{\end{eqnarray}}

\def\diag{{\rm diag \,}}

\def\IF{\relax{\rm I\kern-.18em F}}
\def\II{\relax{\rm I\kern-.18em I}}
\def\IP{\relax{\rm I\kern-.18em P}}

\def\Dsl{\,\raise.15ex\hbox{/}\mkern-13.5mu D} 

\def\IC{\bf C}
\def\IZ{\bf Z}

\def\z2z2{$\IC^3/(\IZ_2\times\IZ_2)$}

\def\id{{\bf 1}}

\def\s{\sigma}

\def\z{\zeta}

\def\bo{{\raise-.3ex\hbox{\large$\Box$}}}               
\def\face{{\raise.2ex\hbox{$\displaystyle \bigodot$}\mskip-2.2mu \llap {$\ddot
        \smile$}}}                                      

\def\leftrightarrowfill{$\mathsurround=0pt \mathord\leftarrow \mkern-6mu
        \cleaders\hbox{$\mkern-2mu \mathord- \mkern-2mu$}\hfill
        \mkern-6mu \mathord\rightarrow$}       
\def\dvec#1{\vbox{\ialign{##\crcr
        \leftrightarrowfill\crcr\noalign{\kern-1pt\nointerlineskip}
        $\hfil\displaystyle{#1}\hfil$\crcr}}}           

\def\beq{\begin{equation}}
\def\eeq{\end{equation}}

\def\beqx{\begin{displaymath}}
\def\eeqx{\end{displaymath}}

\def\beqa{\begin{eqnarray}}
\def\eeqa{\end{eqnarray}}

\begin{document}

\DeclareGraphicsExtensions{.jpg,.pdf,.mps,.png}
\begin{flushright}
\baselineskip=18pt
UPR-1085-T/rev, hep-th/0407178 \\
\end{flushright}

\begin{center}
\vglue 0.8cm

{\Large\bf D6-brane Splitting on Type IIA Orientifolds} \vglue 1cm
{ Mirjam Cveti\v c$^a$, Paul Langacker$^a$, Tianjun Li$^{b}$ and
Tao Liu$^a$} \vglue 0.6cm { $^a$ Department of Physics and
Astronomy,
University of Pennsylvania, \\Philadelphia, PA 19104-6396, USA \\

$^b$ School of Natural Sciences, Institute for Advanced Study,  \\
             Einstein Drive, Princeton, NJ 08540, USA\\}
\end{center}

\thispagestyle{empty}


\begin{abstract}
We study the open-string moduli of supersymmetric D6-branes,
addressing both the string and field theory aspects of D6-brane
splitting on Type IIA orientifolds induced by open-string moduli
Higgsing ($i.e.$, their obtaining VEVs). Specifically, we focus on
the $\IZ_2\times \IZ_2$ orientifolds and address the symmetry
breaking pattern for D6-branes parallel with the orientifold
6-planes as well as those positioned at angles. We demonstrate
that the string theory results, $i.e.$, D6-brane splitting and
relocating in internal space, are in one to one correspondence
with the field theory results associated with the Higgsing of
moduli in the antisymmetric representation of $Sp(2N)$ gauge
symmetry (for branes parallel with orientifold planes) or adjoint
representation of $U(N)$ (for branes at general angles). In
particular, the moduli Higgsing in the open-string sector results
in the change of the gauge structure of D6-branes and thus changes
the chiral spectrum and family number as well. As a by-product, we
provide the new examples of the supersymmetric Standard-like
models with the electroweak sector arising from $Sp(2N)_L\times
Sp(2N)_R$ gauge symmetry; and one four-family example is free of
chiral Standard Model exotics.

\end{abstract}

\newpage

\section{Introduction}
Intersecting D-brane models provide a  framework in which the
particle physics aspects of string theory, such as the origin of
gauge structures and   chiral spectrum, as well as some couplings,
have a beautiful geometric interpretation.

In particular the intersecting D6-brane constructions on
compact Type IIA orientifolds \cite{AN,LU,IB,IB2,LU2,CSU1,CSU2},
wrapping three-cycles of the internal orbifold,
provide a framework where the explicit
conformal field theory techniques in the open string sectors can be employed to
determine the gauge structure, the chiral  spectrum and  (some) couplings.
Within this framework the chiral matter  \cite{douglas} is located at the D6-brane intersection
 points in the internal space.

A large number of non-supersymmetric  three family Standard-like  models
have been constructed, based on the original work \cite{LU,IB,IB2,LU2}  (for a partial list, see
\cite{imr}-\cite{LL}).  These models satisfy the Ramond-Ramond (RR)
tadpole
cancellation conditions; however, since the models are
non-supersymmetric, there are uncancelled
Neveu-Schwarz-Neveu-Schwarz (NS-NS) tadpoles. In addition,
 in these toroidal/orbifold constructions
the string scale is close to the Planck scale, thus
 these models typically suffer from the
large Planck scale corrections at the loop level.

On the other hand, the first supersymmetric three family
Standard-like models with intersecting D6-branes were constructed
in ~\cite{CSU1,CSU2}. These constructions were based on  Type IIA
$T^6/(\IZ_2\times \IZ_2)$ orientifolds. While stable at the string
scale the original constructions possess additional chiral
exotics. (Their phenomenological consequences were studied in
\cite{CLS1,CLS2,CLW,CP2}.) Subsequent work \cite{CP} revealed more
examples of  Standard-like models, as well as systematic
constructions of Grand Unified Georgi-Glashow models \cite{CPS}.
 For  works on the
supersymmetric constructions of  Standard-like models on other orientifolds, see
\cite{Blum,Hon,HO}. Most recently, in
\cite{CLL}   a systematic
search  of $\IZ_2\times \IZ_2$ orientifolds  revealed a number of supersymmetric
 potentially
viable three-family Standard-like models. These constructions are based on the
Pati-Salam gauge  symmetry. The subsequent symmetry breaking via
D-brane splitting  and D-brane recombination gives  no additional
$U(1)$'s near the electroweak
scale, in contrast to the previous supersymmetric
Standard-like Model constructions~\cite{CP} which typically possess additional light neutral
gauge bosons. Generically these models possess the chiral Standard Model exotics,
which however can  be removed from the spectrum  due to the hidden sector
strongly coupled dynamics \cite{CLS1}.

 [There is currently strong activity, in constructions of
 supersymmetric
chiral  solutions with D-branes of  Gepner models;
see \cite{Blum2,Ilke,BW,schell} and references therein.
While there seems to be \cite{schell}
a large class of  three-family Standard-like Models with no chiral exotics, these
exact conformal field theory models are located at the special points in moduli
space where the geometric picture is lost, and the couplings, such as Yukawa couplings, do
not possess hierarchies associated with the size of the internal spaces, such
as in the case of the toroidal orbifolds with intersecting D-branes.]

The main focus of the previous constructions based on intersecting D6-branes on
compact orbifold  constructions  was  on obtaining   the  gauge
symmetry and the chiral spectrum  of the effective theory at the
string scale. In particular, further study is needed to shed light
on the gauge symmetry breaking patterns, both from the point of
view of deforming the original string construction, {\it i.e.},
 by employing the splitting  or recombination of D-branes, and in the dual field
 theory  approach. The latter involves  giving vacuum expectation values (VEVs) to chiral
 superfields-moduli, associated with the brane deformation, referred to as
  Higgsing.
The intriguing property of the  Type IIA orientifold constructions with the
intersecting D6-branes is that  the gauge symmetry breaking patterns have a
geometric interpretation either in terms of recombination of branes that wrap
different intersecting cycles, or in terms of parallel splitting of branes that wrap
parallel  (non-rigid) cycles. [For related work, see, e.g.,
\cite{LUS,EM} and for a review,  \cite{AS}.]

The focus of this paper will be on the second
phenomenon of parallel splitting of branes. This is a specific phenomenon due to
the fact that for toroidal orbifolds the cycles wrapped by D-branes are not
rigid and thus the continuous parallel splitting of branes  (consistent with
preservation of supersymmetry)  can take place.  In the  T-dual picture this
geometric deformation can be interpreted as the introduction of Wilson lines in the
open string sector of string theory. [For an analogous  explicit  study
of continuous Wilson lines for (fractional) D9-branes, their
 T-dual  interpretation as moving D3-branes,
 as well as the  field theory analysis of Higgsing for a specific supersymmetric
 Type IIB chiral model on  compact
 $\IZ_3$ orientifold \cite{ABS}, see \cite{CL}.]

The purpose of this paper is a few fold. First we would like to
gain insight into the detailed interpretation of the symmetry
breaking patterns in the case of splitting of branes that are
parallel with orientifold planes. In particular, the breaking
patterns in this case are non-trivial and also result in the
change of the chiral spectrum and number of families.
Specifically, we discuss in detail the string theory results for
all such breaking patterns in the case of
 toroidal $\IZ_2\times \IZ_2$
 orbifolds  with the six-torus factorized as a product of three two-tori.
  The breaking pattern is qualitatively different when
 branes split from the orientifold planes
in one two-torus, two two-tori and  three  two-tori directions, respectively.
The gauge symmetry breaking pattern typically involves $Sp(2N)$ groups.
We also discuss the symmetry breaking pattern
when the branes are at angles relative
 to the orientifold planes; there   the symmetry breaking is that of $U(N)$.

We also demonstrate  how this string theoretic interpretation of
symmetry breaking patterns is in one to one  correspondence with
the field  theoretical interpretation in terms of Higgsing of
open-string moduli living on the branes. For the branes parallel
with the orientifold planes  the splitting in one, two and three
two-tori corresponds to   Higgsing of  one, two and three chiral
superfields in the anti-symmetric representation of $Sp(2N)$,
respectively. For the branes at angles the Higgsing takes place
due to the one, two and three moduli in the adjoint representation
of $U(N)$, respectively.

As a by-product of this analysis we present three Standard-like
models, two supersymmetric ones and one non-supersymmetric one, in
which the electroweak symmetry part of the Standard model is
associated with the branes parallel to the orientifold planes. As
a global supersymmetric construction this possibility has not been
addressed, yet.  In particular, the four-family supersymmetric
example and the three-family non-supersymmetric example are free
from massless Standard Model chiral exotics.

The paper is organized as follows. In Section 2 we
discuss the string theory  aspects of D-brane splitting  and the
resulting symmetry breaking patterns for the Type IIA $\IZ_2\times
\IZ_2$ orientifold, both for D6-branes parallel (Section 2a) and
not parallel (Section 2b) to O6-planes. In section 3 we analyze
its dual field theory   interpretation in terms of Higgsing, again
both for D6-branes parallel (Section 3a) and not parallel (Section
3b) to O6-planes. In section 4, we present the new Standard-Model-like
models  in which the electroweak part is generated by splitting of
branes parallel with the orientifold planes. In Section 4a a
four-family model without Standard Model chiral exotics is given.
 In Sections 4b a non-supersymmetric three family example is presented.
 The Higgsing can occur at a low or intermediate scale, which however
 must be large enough to suppress new flavor changing neutral
 current interactions.
In Section 4c we  present a consistent supersymmetric   three-family
Standard-like model based on $Sp(2)_L\times Sp(2)_R$ electroweak
symmetry. The Standard Model part of this  model was originally
proposed in \cite{cim5,cim6} on a toroidal orientifold;
 it  was  locally supersymmetric but did not cancel RR tadopoles.
   Conclusions and open
questions are given in Section 5.

\section{D6-brane Splitting  in String Theory}

The intriguing property of  type IIA orientifolds with
intersecting D6-branes is the geometric interpretation both of the
gauge symmetry and of the origin of the chiral spectrum. In
particular, different gauge group factors are generated from
different D6-brane stacks/D-brane configurations, wrapping
three-cycles of the orbifolds. The concrete group structure is
determined by the orientifold and orbifold projections on the
Chan-Paton factors for gauge bosons in the open string sector that
start and end on the same stack of D-branes. The resulting
symmetry breaking pattern
 is determined by the number of D-branes and the specific
  discrete projection, associated with the orientifold and/or orbifold
  symmetries,   on Chan-Paton indices.

  The second intriguing feature is that the symmetry
breaking pattern associated with  stacks of D6-branes corresponds
to a specific  geometric operation  on D6-branes. A first such
typical process involves  D6-branes from different (intersecting)
stacks  recombining and  forming a new brane configuration.
(Within the specific construction based on the $\IZ_2\times \IZ_2$
orientifold,  the implications of  D6-brane recombination for the
gauge symmetry breaking pattern and the change in the chiral
spectrum have been discussed  \cite{CSU2}.). It can take place for
intersecting D-brane stacks, where two group factors are broken
down to one and the Higgs role is played by the massless superfields in
bi-fundamental representation located at the   intersection of
 two brane stacks. In the non-supersymmetric case, D-brane
recombination can be interpreted as a tachyon condensation
process.

 The second process
involves a   stack of D6-branes that split parallel to each other.
Comparing with D-brane recombination, D-brane splitting involves
one single D-brane stack, and a single  group factor is broken
down to subgroups. For D6-branes parallel (non-parallel) with the
orientifold planes the gauge group factor is $Sp(2N)$  ($U(N)$),
and in the dual field theory the Higgs particles-moduli are three
$N=1$ chiral superfields in the antisymmetric (adjoint)
representations of the respective gauge group factors.
 The  three  copies of the Higgs fields are in one to one correspondence with
  the geometric interpretation of
 splitting branes in one, two and three two-tori, which are generic deformations
 of the factorizable three-cycles wrapped by D6-branes on toroidal orbifolds.

 For the sake of concreteness, in
order to illustrate the correspondence between the D-brane splitting
and field theory Higgsing in type IIA orientifold constructions,
we focus on the
  ${\bf T}^6 /(\IZ_2 \times \IZ_2)$ orientifold.

We briefly review  the construction of the Type IIA  ${\bf T}^6
/(\IZ_2 \times \IZ_2)$ orientifold (for more details, refer to
\cite{CSU2}). We consider ${\bf T}^{6}$ to be a six-torus
factorized as ${\bf T}^{6} = {\bf T}^{2} \times{\bf T}^{2} \times
{\bf T}^{2}$ whose complex coordinates are $z_i$, $i=1,\; 2,\; 3$
for the $i$-th two-torus, respectively. The $\theta$ and $\omega$
generators for the orbifold group $\IZ_{2} \times \IZ_{2}$, which
are associated with their twist vectors $(1/2,-1/2,0)$ and
$(0,1/2,-1/2)$, respectively, act on the complex coordinates of
${\bf T}^6$ as
\begin{eqnarray}
& \theta: & (z_1,z_2,z_3) \to (-z_1,-z_2,z_3)~,~ \nonumber \\
& \omega: & (z_1,z_2,z_3) \to (z_1,-z_2,-z_3)~.~\,
\label{orbifold}
\end{eqnarray}
When a specific brane configuration is invariant under
these orbifold actions, the corresponding Chan-Paton factors are subject to
their  projections, as discussed in the following Subsections.
 [The fact that  D6-branes are invariant under orbifold
projections does not imply  that their intersection points will be.
 The final spectrum, however, turns out to be rather insensitive
to this subtlety in the case of the ${\bf T}^6 /(\IZ_2 \times
\IZ_2)$ orientifold construction. See  \cite{CSU2} for further
discussions.]

The orientifold projection is implemented by
gauging the symmetry $\Omega R$, where $\Omega$ is world-sheet
parity, and $R$ acts as
\begin{eqnarray}
 R: (z_1,z_2,z_3) \to ({\ov z}_1,{\ov z}_2,{\ov
z}_3)~.~\, \label{orientifold}
\end{eqnarray}
There are four kinds of orientifold 6-planes (O6-planes) due to the
action of $\Omega R$, $\Omega R\theta$, $\Omega R \omega$,
and $\Omega R\theta\omega$, respectively. Their configurations
have been tabulated in Table \ref{orientifold table} and presented
geometrically for rectangular two-tori in  Figure \ref{oplanes}.
In the following we shall focus on the case  of rectangular tori, only.

\subsection{D6-Branes Parallel with O6-Planes}
String states in the open string sector are  subject to orbifold and orientifold
projections on their Chan-Paton indices.  For the states associated with the branes
 located on the top of the orientifold planes,  the brane configuration is
invariant under the action of  the orientifold  $\Omega\,  R$
projection as well as  the orbifold projections $\omega$ and
$\theta$. Thus, the Chan-Paton indices for the open-string states
in this sector are subject to all three projections.  We choose
the following specific $\gamma$ representations for these group
elements acting  on  the Chan-Paton indices associated with the
stack of $4N$  on the top of the orientifold fixed plane:
\begin{eqnarray}
\gamma_{\theta } & = &
\diag(i \id_{N},-i \id_{N}\, ;
-i \id_{N},i \id_{N})\, ,  \nonumber \\
\gamma_{\omega} & = & \diag \left[ \pmatrix{0 & \id_{N} \cr
-\id_{N} & 0 } \; ; \;
\pmatrix{0 & \id_{N} \cr -\id_{N} & 0 } \right]\, ,  \nonumber \\
\gamma_{\Omega R} & = & \pmatrix{  0& \id_{2N}  \cr \id_{2N}& 0 \cr
 }\, .\label{gamma}
\end{eqnarray}
The  above representations  are the same as those in \cite{CSU2}.
The  actions for the orbifold groups form a projective
representation as explained in \cite{CSU2}.
 The $\gamma$ representations are   compatible with  all the  symmetry
constraints on the orientifold group  action. They satisfy the
following conditions:
\begin{equation}
\gamma_{\theta}^2=-\id_{4N}\, , \ \ \ \gamma_{
\omega}^2=-\id_{4N}\, ,
\end{equation}
 and
\begin{equation}
(\gamma_{\theta})^*= \gamma_{\Omega  R}\gamma_{\theta} \gamma_{\Omega
R}\, , \ \ \
\\(\gamma_{\omega})^*= \gamma_{\Omega  R}\gamma_{\omega} \gamma_{\Omega
R}\, ,
\end{equation}
where $\gamma_{\Omega
R}$ is chosen to be real,  which is consistent with the choice for the
orientifold projection on $D9$-branes in Type IIB string theory. (See, e.g.,  \cite{cpw}
for details.) In addition, $\gamma_\theta$ and $\gamma_\omega$ commute, {\it i.e.},
there is no
discrete torsion. Note  also that $\hbox{Tr}\, (\gamma_\theta)= \hbox{Tr}\, (\gamma_\omega ) ={
0}$, {\it i.e.}, there are no twisted tadpoles.

Since the configuration of $4N$ D6-branes positioned on the top of the
O6-plane is invariant under all orbifold and orientifold
projections, the Chan-Paton indices   $\lambda$ ($4N\times 4N$
matrix), associated with the gauge boson states, should be
invariant under all these projections and should satisfy:
 \begin{eqnarray}
 \lambda&=& -\gamma_{\Omega R}\, \lambda^T \, \gamma_{\Omega
R}^{-1},\nonumber\\
 \lambda&=& \gamma_\theta \,\lambda  \, \gamma_\theta^{-1},\nonumber\\
 \lambda&=& \gamma_\omega\, \lambda \, \gamma_\omega^{-1}.
 \end{eqnarray}
This sequence of projections  yields the following
symmetry breaking chain:
\begin{eqnarray}
U(4N) &\to& \hbox{(due to} \ \gamma_\theta ) \ U(2N)\times
U(2N) \nonumber \\
&\to& \hbox{(due  to }\ \gamma_\omega ) \ U(2N) \nonumber
\\
&\to& \hbox{(due to} \  \gamma_{\Omega R})\ Sp(2N)~. ~\,
\end{eqnarray}
Therefore, the final gauge symmetry is $Sp(2N)$ with the
 resulting  $\lambda$ matrix of the form:
\begin{eqnarray}
\lambda  &=&  \pmatrix{ A & 0     & 0        & B  \cr
                        0    &A &-B    & 0        \cr
                        0    &C &-A^T & 0         \cr
                      -C& 0     & 0        & -A^T \cr
 }\, ,
\end{eqnarray}
where  $A$ is an arbitrary $N \times  N$ matrix,  and $B$ and  $C$
are $N   \times N$ symmetric matrices. The structure of the
Chan-Paton indices thus confirms that this is the adjoint
representation of the
  $Sp(2N)$ gauge symmetry.

\begin{figure}
\begin{center}
\scalebox{0.55}{{\includegraphics{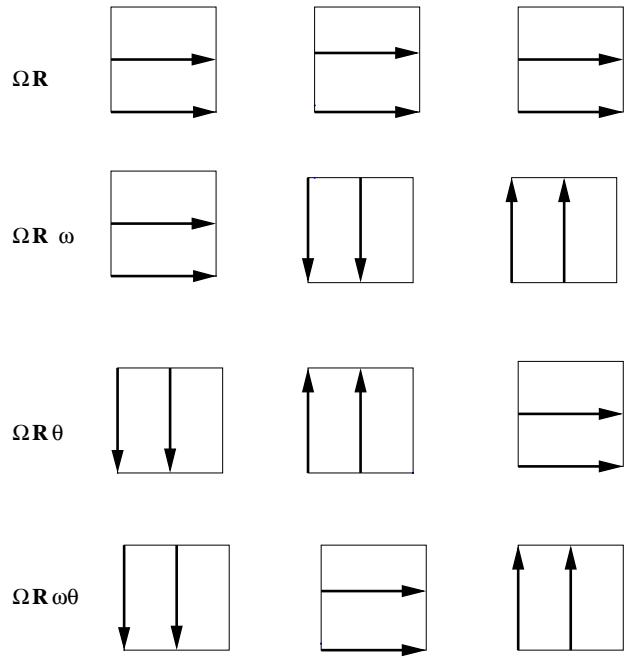}}}
\end{center}
\caption[]{\small
The locations of O6-planes fixed under the orientifold actions
$\Omega R$,  $\Omega R \omega$,  $\Omega R \theta$, and $\Omega R
\omega \theta$
 (denoted by  bold solid lines)
for the case of a six-torus factorized on  three rectangular
two-tori.  }
%
\label{oplanes}
\end{figure}

 Since there are a number of the fixed O6-planes
 (see Figure \ref{oplanes}),  one can position sets of $4N_i$ D6-branes on
 different fixed O6-planes, and then obtain in general $\prod_i \, Sp(2N_i)$
 gauge symmetry. (Of course the final set of brane configurations has
to cancel  the R-R tadpoles, which puts
constraints  on the allowed values of $N_i$.)


\renewcommand{\arraystretch}{1.4}
\begin{table}[t]
\caption{Wrapping numbers of the four O6-planes. $\beta_i$ is
equal to 0 and 1 for rectangular and tilted two-tori,
respectively.} \vspace{0.4cm}
\begin{center}
\begin{tabular}{|c|c|c|}
\hline
  Orientifold Action & O6-Plane & $(n^1,l^1)\times (n^2,l^2)\times
(n^3,l^3)$\\
\hline
    $\Omega R$& 1 & $(2^{\beta_1},0)\times (2^{\beta_2},0)\times
(2^{\beta_3},0)$ \\
\hline
    $\Omega R\omega$& 2& $(2^{\beta_1},0)\times (0,-2^{\beta_2})\times
(0,2^{\beta_3})$ \\
\hline
    $\Omega R\theta\omega$& 3 & $(0,-2^{\beta_1})\times
(2^{\beta_2},0)\times
(0,2^{\beta_3})$ \\
\hline
    $\Omega R\theta$& 4 & $(0,-2^{\beta_1})\times (0,2^{\beta_2})\times
    (2^{\beta_3},0)$ \\
\hline
\end{tabular}
\end{center}
\label{orientifold table}
\end{table}

 In  the following we shall address the gauge
  symmetry breaking pattern when sets of D6-branes split parallel with
O6-planes.

\medskip
{\bf D-Branes splitting in one two-torus}
\medskip

\begin{figure}
\begin{center}
\scalebox{0.65}{{\includegraphics{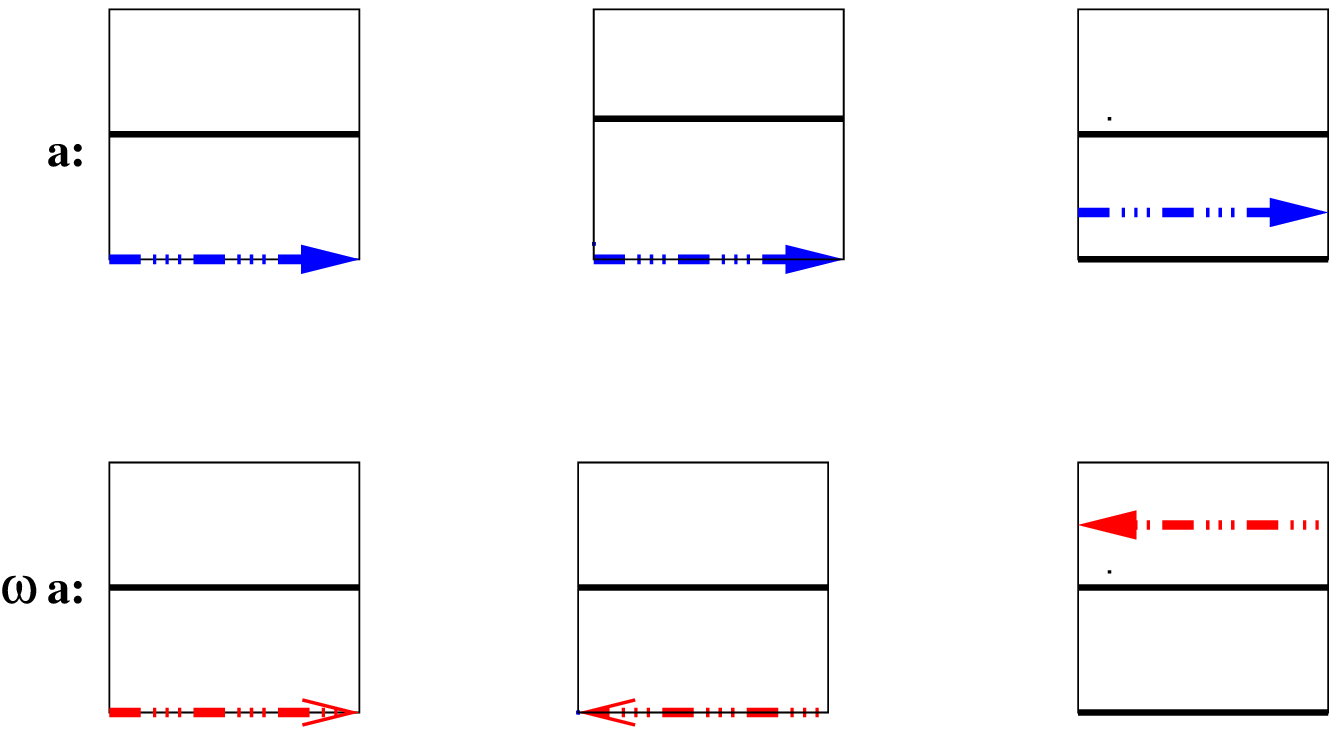}}}
\end{center}
\caption[]{\small  Two distinct  brane configurations (denoted by dash-dotted
lines): configuration $a$ and its $\omega$ image
in the case of  D6-branes being split away
from the orientifold plane in one, say, third, two-torus. Bold solid lines
denote the orientifold planes. }
\label{12torus}
\end{figure}
When one moves branes away from the O6-plane in only one
two-torus direction, $e.g.$, the third one, there are two distinct
configurations: $a$ and $\omega a$ (see Figure \ref{12torus}). However, these configurations
are invariant under the two remaining Chan-Paton projections,
$i.e.$, $\gamma_{\Omega R \omega}$ and $\gamma_{\theta}$, which
yield the symmetry breaking pattern with  $Sp$ gauge groups.
Specifically,   if one takes
$2(2k_i)$-multiples of D6-branes, the
symmetry breaking chain is:
\begin{eqnarray}
Sp(2N)
 &\to&  \hbox{(due to branes on two images)}\,
Sp(2N-2k_i)\times U(2k_i)\,
 \nonumber\\
&\to&
 \hbox{(due to}\  \gamma_{\Omega R  \omega} \hbox{ and } \gamma_\theta \,
)
 \  Sp(2N-2k_i)\times Sp(2k_i) ~.~\,
\label{USp breaking pattern1}
\end{eqnarray}
More explicitly, the Chan-Paton factor $\lambda$ for the  gauge bosons
associated with the $2(2k_i)$  D6-branes moved in one two-torus direction is
subject  to  the following projections:
\begin{eqnarray}
 \lambda&=& -\gamma_{\Omega R}\, \gamma_\omega \, \lambda^T  \, \gamma_\omega
 ^{-1}\, \gamma_{\Omega R}^{-1},\nonumber\\
 \lambda&=& \gamma_\theta\,  \lambda \,  \gamma_\theta^{-1},
 \end{eqnarray}
and  therefore the final form of $\lambda$ is:
\begin{eqnarray}
\lambda  &=&  \pmatrix{ A_1 & 0     & 0        & B_1  \cr
                         0    &A_2 &B_2    & 0        \cr
                        0    &C_2 &-A_2^T & 0         \cr
                       C_1& 0     & 0        & -A_1^T \cr
 }\, ,\label{cp12t}
\end{eqnarray}
where  $A_1$ and $A_2$ are  arbitrary $k_i \times  k_i$
matrices, and $B_1$, $B_2$, $C_1$ and $C_2$ are $k_i
\times k_i$ symmetric matrices.  This is precisely the  Chan-Paton  matrix,
which can be cast after  a suitable interchange of rows and columns  in a
block-diagonal form.  This structure is
 associated with the gauge symmetry of configuration $a$, say,  the matrices
with  subscript 1, for the adjoint representation of $Sp(2k_i)$,
 and the  $\omega \, a$  image,  say, the matrices with subscript 2, also for
 the  adjoint representation of $Sp(2k_i)$.
 The action of  $\gamma_\omega$  on the above Chan-Paton matrix
  precisely interchanges matrices with index 1 and 2: $A_1\to A_2$,
 $B_1\to -B_2$ and $C_1\to -C_2$, and
then effectively maps the Chan-Paton indices of
 configuration $a$ to those of  the image $\omega\, a$.
 (see Figure \ref{12torus}).
 Thus, the resulting  gauge group is $Sp(2k_i)$.

The above result is unique. Even though we started with the general
$\lambda$, the  $\gamma_\theta$ and $\gamma_{\Omega\, R\, \omega}$ produced the
form in Eq. (\ref{cp12t}) that can be cast in  a block-diagonal form  associated with
the  gauge structure of the  $a$ and $\omega\, a$ configurations.
The result is the same with the more restricted   Ansatz:
\begin{eqnarray}
\lambda  &=&  \pmatrix{ A_1 & 0     & 0        & B_1  \cr
                         0    &A_2 &B_2    & 0        \cr
                        0    &C_2 & D_2 & 0         \cr
                       C_1& 0     & 0        & D_1 \cr
 }\, ,\label{cp12t0}
\end{eqnarray}
where $A_{1,2}, \ B_{1,2}, \ C_{1,2}, \ D_{1,2}$ are general $N\times N$
matrices.  This is compatible with the action
of $\gamma_\omega$, {\it i.e.}, it interchanges the matrices with index 1 and 2 associated
with $a$ and $\omega a$ configurations, respectively.
Before the  $\gamma_\theta$ and  $\gamma_{\Omega R  \omega}$ projections
 the gauge group is
$U(2k_i)$,  associated with  $2k_i$ branes for each of the two configurations.

\medskip
{\bf D-Branes splitting in two two-tori}
\medskip

\begin{figure}
\begin{center}
\scalebox{0.65}{{\includegraphics{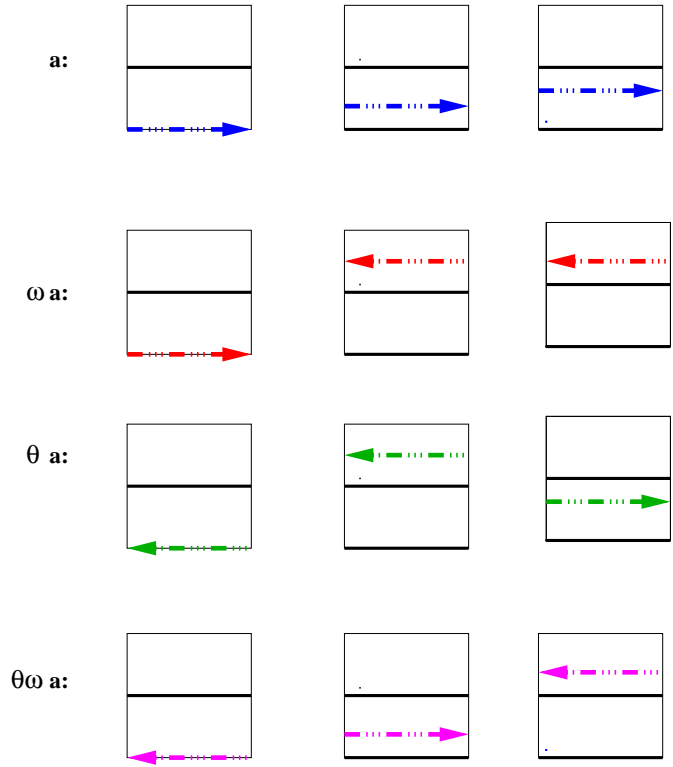}}}
\end{center}
\caption[]{\small  Four distinct  brane configurations: $a$ and its $\omega$ ,
$\theta$ and $\theta\, \omega$ images
in the case of  D6-branes being split away
from the orientifold plane in two two-tori directions, say, in the second and
third two-torus. }
\label{22tori}
\end{figure}

When one moves branes away from the O6-plane in two two-tori
directions, there are four distinct configurations: $a$, $\theta
a$, $\omega a$ and $\theta \omega a$ (see Figure \ref{22tori}).
The Chan-Paton matrix that reflects the  orbifold action on these four
configurations can be cast in the form:
\begin{eqnarray}
\lambda  &=&  \pmatrix{0   & A_1 & 0  & 0   & 0   & 0   & B_1 &  0  \cr
                       A_2 &  0  & 0  & 0   & 0   & 0   & 0   & B_2 \cr
                       0   &  0  & 0  & A_3 & B_3 & 0   & 0   & 0   \cr
                       0   &  0  &A_4 & 0   & 0   & B_4 & 0   & 0   \cr
               0   &  0  &C_4 & 0   & 0   & D_4 & 0   & 0   \cr
                       0   &  0  & 0  & C_3 & D_3 & 0   & 0   & 0   \cr
                       C_2 &  0  & 0  & 0   & 0   & 0   & 0   & D_2 \cr
                       0   &C_1  & 0  & 0   & 0   & 0   & D_1 & 0   \cr
 }\, ,\label{cp22t0}
\end{eqnarray}
or
\begin{eqnarray}
\lambda  &=&  \pmatrix{A_1   & 0 & 0  & 0   & 0   & 0   & B_1 &  0  \cr
                       0 &  A_2  & 0  & 0   & 0   & 0   & 0   & B_2 \cr
                       0   &  0 & A_3 & 0 & B_3 & 0   & 0   & 0   \cr
                       0   &  0  & 0  & A_4   & 0   & B_4 & 0   & 0   \cr
               0   &  0  &C_3 & 0   & D_3   & 0 & 0   & 0   \cr
                       0   &  0  & 0  & C_4 & 0 & D_4   & 0   & 0   \cr
                       C_1 &  0  & 0  & 0   & 0   & 0   & D_1   & 0 \cr
                       0   &C_2  & 0  & 0   & 0   & 0   & 0 & D_2   \cr
 }\, ,\label{Ncp22t0}
\end{eqnarray}
where $A_{1,2,3,4}$,  $B_{1,2,3,4}$, $C_{1,2,3,4}$, and $D_{1,2,3,4}$ are
general $k_i\times k_i$ matrices.  The gauge group is $U(2k_i)$ associated with
$(2k_i)$ branes on each of the  four configurations.
{ The $8\times 8$ matrix $\lambda$, whose entries are $k_i\times k_i$ matrices,
can be considered as a $4\times 4$ matrix,
where all the entries are $2\times 2$ matrices
in which each entry is still a $k_i\times k_i$ matrix.
 These $2\times 2$ matrices like in Eq. (\ref{Ncp22t0})
 must be diagonal because the off-diagonal components
correspond to the open string states which stretch between the different
stacks (images) of D6-branes and then are massive.} However,
 there is still one remaining nontrivial projection on
Chan-Paton indices, say, $\gamma_{\Omega R \omega}$ under which
these configurations are invariant if branes are moved away in the
second and third two-tori directions (see Figure \ref{22tori}).
In this case  the $\gamma_{\Omega R
\omega}$  projection (with representation in Eq. (\ref{gamma})) acts separately on the
matrices with indices $1$, $2$, $3$ and $4$,  yielding:
\begin{eqnarray}
\lambda  &=&  \pmatrix{0   & A_1 & 0  & 0   & 0   & 0   & B_1 &  0  \cr
                       A_2 &  0  & 0  & 0   & 0   & 0   & 0   & B_2 \cr
                       0   &  0  & 0  & A_3 & B_3 & 0   & 0   & 0   \cr
                       0   &  0  &A_4 & 0   & 0   & B_4 & 0   & 0   \cr
               0   &  0  &C_4 & 0   & 0   & -A_4^T & 0   & 0   \cr
                       0   &  0  & 0  & C_3 & -A_3^T & 0   & 0   & 0   \cr
                       C_2 &  0  & 0  & 0   & 0   & 0   & 0   & -A_2^T\cr
                       0   &C_1  & 0  & 0   & 0   & 0   & -A_1^T & 0   \cr
 }\, ,\label{cp22t}
\end{eqnarray}
or
\begin{eqnarray}
\lambda  &=&  \pmatrix{A_1  & 0 & 0  & 0   & 0   & 0   & B_1 &  0  \cr
                       0 &  A_2  & 0  & 0   & 0   & 0   & 0   & B_2 \cr
                       0   &  0  & A_3  & 0 & B_3 & 0   & 0   & 0   \cr
                       0   &  0  & 0 & A_4   & 0   & B_4 & 0   & 0   \cr
               0   &  0  &C_3 & 0   &  -A_3^T  & 0 & 0   & 0   \cr
                       0   &  0  & 0  & C_4 & 0 &  -A_4^T  & 0   & 0   \cr
                       C_1 &  0  & 0  & 0   & 0   & 0   &  -A_1^T  & 0 \cr
                       0   &C_2  & 0  & 0   & 0   & 0   & 0 &  -A_2^T  \cr
 }\, ,\label{Ncp22t}
\end{eqnarray}
where  $A_{1,2,3,4}$ are general $k_i\times k_i$ matrices, and $C_{1,2,3,4}$ and
$B_{1,2,3,4}$ are symmetric $k_i\times k_i$ matrices.
Therefore, the gauge symmetry is $Sp(2k_i)$ with  the
 Chan-Paton matrix associated with
all four  configurations  $a$, $\omega a$, $\theta a$ and
$\theta \omega a$.

In this process, one
 takes $2(4k_i)$-multiples of branes and the breaking  pattern is:
\begin{eqnarray}
Sp(2N)&\to& \hbox{(due to branes on four images)}\
Sp(2N-4k_i)\times U(2k_i) \nonumber \\
&\to& \hbox{(due to}\ \gamma_{ \Omega R \omega}  )\
Sp(2N-4k_i)\times Sp(2k_i)~.~\,
\label{USp breaking pattern2}
\end{eqnarray}

\medskip
{\bf D-Branes splitting in three two-tori}
\medskip
\begin{figure}
\begin{center}
\scalebox{0.65}{{\includegraphics{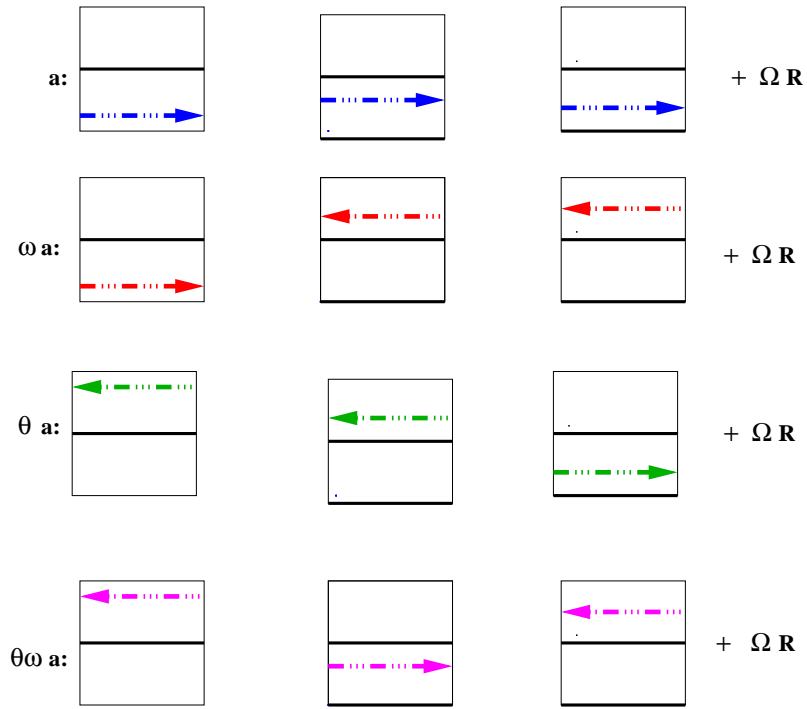}}}
\end{center}
\caption[]{\small  Four  distinct  brane configurations: $a$
configuration, its $\omega$ , $\theta$ and $\theta\omega$ images
as well as four $\Omega \, R$ (not depicted on the figure) in the
case of  D6-branes being split away from the orientifold plane in
all three two-tori directions.} \label{32tori}
\end{figure}

If one splits D6-branes in all three toroidal directions,
there are now eight distinct configurations: $a$, $\theta\,  a$, $\omega\,  a$,
${\omega \, \theta \,  a} $, and four new images due to the $\Omega\,
R$ action (see Figure \ref{32tori}). Thus one has
 to move away from the O6-planes at least eight
multiples of branes. Since  there is no further projection on Chan-Paton
indices, moving sets of $2(4k_i)$ branes away from O6-planes in all three
toroidal directions results in the symmetry breaking pattern:
\begin{equation}
Sp(2N)\to  \hbox{(due to branes on eight images)}\ Sp(2N-4k_i)\times U(k_i) \label{USp breaking
pattern3}~.~\,
\end{equation}

In general, if we split $2(2k_i^1)$ D6-branes in any one of the three two-tori,
$2(4k_j^2)$ D6-branes in any two of the three two-tori, and
$2(4k_l^3)$ D6-branes in three two-tori, the gauge symmetry breaking
pattern is:
\begin{eqnarray}
Sp(2N) &\to&   Sp(2N-\sum_i 2 k^1_i -\sum_j 4 k^2_j -\sum_l 4 k_l^3 )
\times \prod_i Sp(2 k^1_i) \nonumber \\
&& \times \prod_j Sp(2 k^2_j) \times
 \prod_l  U(k_l^3) \label{General USp breaking
pattern}~.~\,
\end{eqnarray}

\begin{figure}
\begin{center}
\scalebox{0.6}{{\includegraphics{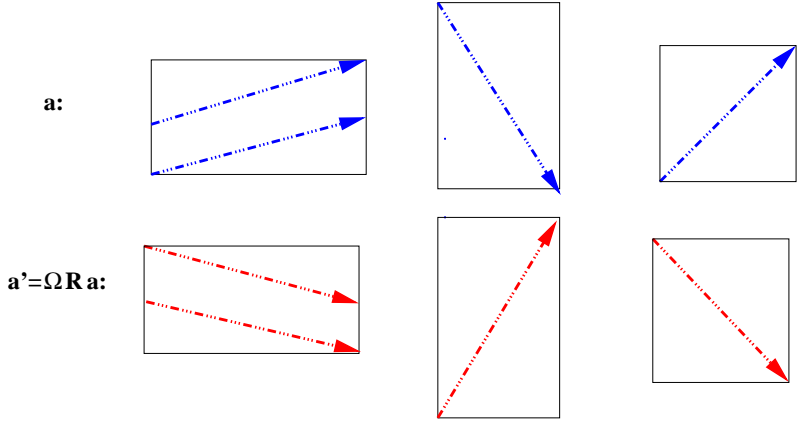}}}
\end{center}
\caption[]{\small  The D6-brane configuration (dash-dotted line) wrapping
a (supersymmetric)  three-cycle
invariant under orbifold projection. The second configuration is its orientifold
image. }
\label{angle}
\end{figure}

\subsection{D6-Branes not Parallel to O6-Planes}

For completeness let us also  mention the splitting of D6-branes in stacks that
are not parallel to the O6-planes. In particular, one chooses first
 $2(2N)$ D6-branes,  wrapping  a factorizable  three-cycle that is invariant
 under the $\IZ_2\times \IZ_2$ action. In this case $2N$
D6-branes wrap  the original three-cycle, {\it i.e.}, configuration
$a$,
 and $2N$ branes wrap a three-cycle that is
its $\Omega \, R$ image, {\it i.e.},
 the  configuration  $a'\equiv \Omega\, R\, a$ (see
Figure \ref{angle}).  The Chan-Paton matrix can therefore be cast in a block
diagonal from:
\begin{equation}
\lambda=\diag(A\, ;
D)\, ,
\end{equation}
where $A$ and $D$ are  general $2N\times 2N$ matrices, associated with the $a$ and
$a'$ configurations, respectively. Since the
$\gamma_\theta$ and $\gamma_\omega$ matrices in Eq. (\ref{gamma}) are
also diagonal, it is sufficient to do projections only on the $A$ matrix.  (The
action on $D$  goes in parallel and just reflects the fact that for each  $a$
there is  an $a'$ image.)
 Since the $a$  configuration is invariant
under $\theta$ and $\omega$ actions, one has to perform the
$\gamma_\theta$ and $\gamma_\omega$ projections on the Chan-Paton
indices, yielding the $U(N)$ group factor \cite{CSU2}.

\medskip
{\bf D-Branes splitting in one two-torus}
\medskip
\begin{figure}
\begin{center}
\scalebox{0.6}{{\includegraphics{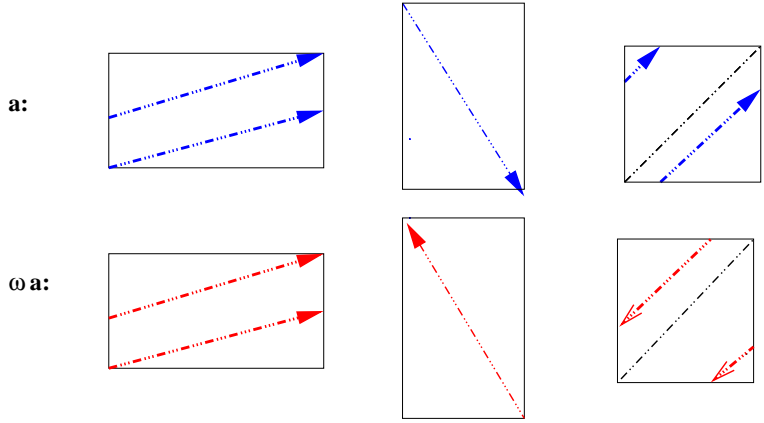}}}
\end{center}
\caption[]{\small  The D6-brane configuration  (dash-dotted line)
wrapping a (supersymmetric) 3-cycle which is  not invariant under
the orbifold projection, say in the third two-torus.  Its $\omega$
image  is therefore a different configuration.
  }
\label{1angle}
\end{figure}
\begin{figure}
\begin{center}
\scalebox{0.6}{{\includegraphics{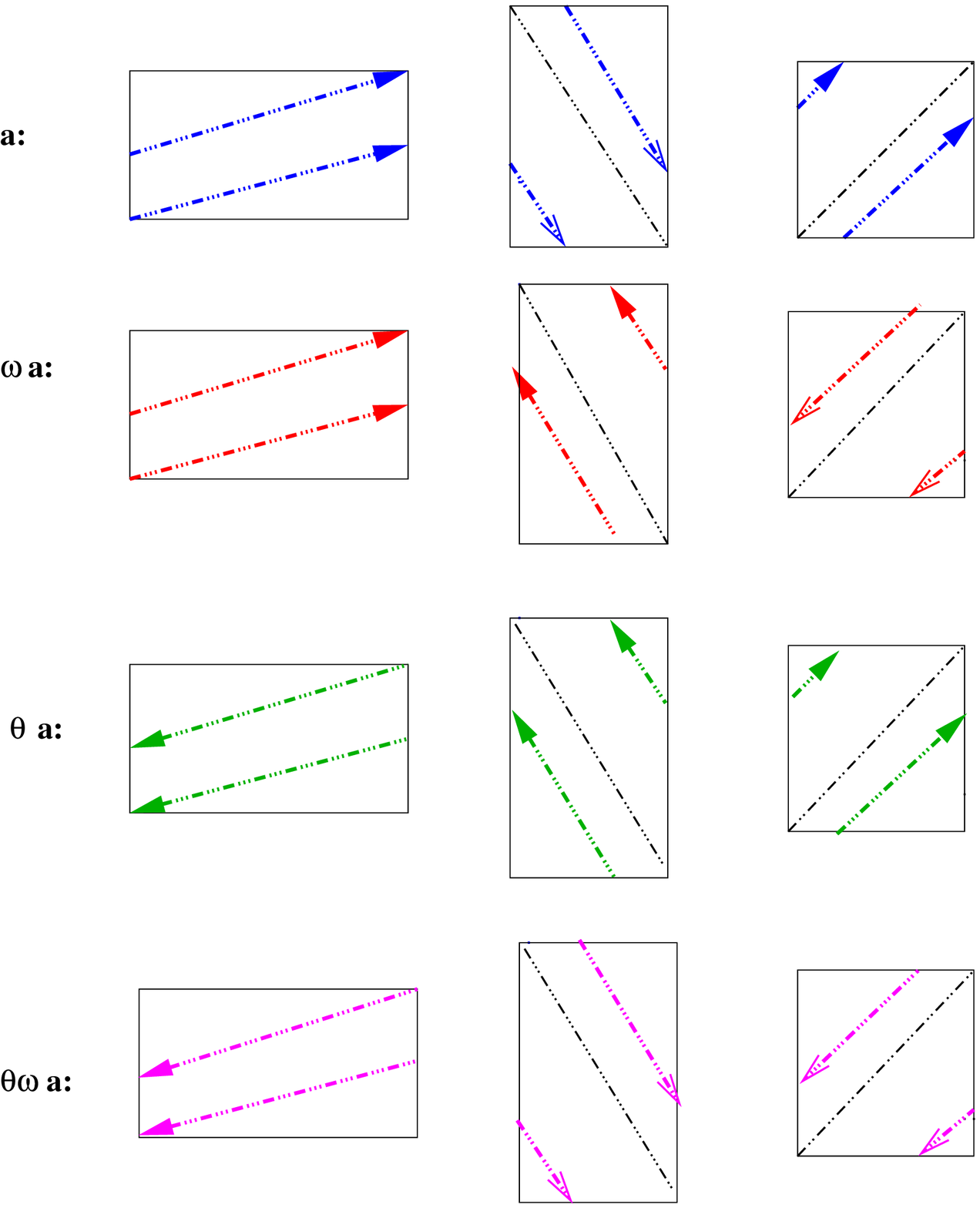}}}
\end{center}
\caption[]{\small  The D6-brane configuration  (dash-dotted line)
wrapping a (supersymmetric) 3-cycle which is not invariant under
the orbifold projection, say in the  second and third
two-tori. Its three orbifold images are therefore different
configurations.
    }
\label{2angle}
\end{figure}

When one moves $2(2k_i)$ branes in one two-torus direction,
{\it e.g.}, in the third two-torus direction, there is now the
$a$-configuration and its $\omega \ a$ image (in addition to
the $a'$-configuration and $\omega\  a'$ image), see Figure \ref{1angle}.
It is sufficient to perform  the $\theta$ projection on the Chan-Paton
matrix $\lambda$, which is a $2k_i\times 2k_i$  matrix of the form:
 \begin{eqnarray}
\lambda  &=&  \pmatrix{ A_1      & 0  \cr
                        0 & A_2 \cr
 }\, , \label{cp1u}
\end{eqnarray}
where $A_1$ and $A_2$ are arbitrary $k_i\times k_i$ matrices.
 This restricted form
of the matrix is compatible with the action of $\gamma_\omega$: $A_1\to A_2$,
mapping the Chan-Paton indices of  the $a$ configuration to
those of $\omega \, a$. (Actions on
orientifold images $D_1$ and $D_2$ of course go in paprallel.)
So, the gauge group is $U(k_i)$.

The representation of $\gamma_\theta$ is only the upper half of the matrix
$\gamma_\theta$ in Eq. (\ref{gamma}) because we neglect the
$\Omega \, R$ images. Then
the matrix in Eq. (\ref{cp1u}) is invariant under  $\gamma_\theta$ projection.
Therefore, the  gauge symmetry in this case is still
$U(k_i)$ and the breaking pattern
is of the form:
\begin{eqnarray}
U(N) &\to&  \hbox{(due to branes on two images)}\,
U(N-k_i)\times U(k_i)\,  \nonumber \\
 &\to & (\hbox{due to} \ \gamma_\theta\ ) \ U(N -k_i)\, \times U(k_i)
\, .
\end{eqnarray}

\medskip
{\bf D-Branes splitting in two two-tori}
\medskip

When one moves branes in two two-tori directions, there are four
distinct configurations: $a$, $\theta \, a$, $\omega \,  a$ and
$\theta\,  \omega\, a$ (and of course  four distinct $\Omega R$
images), see Figure \ref{2angle}. In this case, one has to take
$2(4k_i)$ branes  and put them on all these distinct
configurations (there are  no  projections on Chan-Paton indices).
The symmetry breaking pattern is:
\begin{equation}
U(N)
 \to  \hbox{(due to branes on four images)}\,
U(N-2k_i)\times U(k_i)\, .
 \label{Utwo}
\end{equation}

\medskip
{\bf D-Branes splitting in three  two-tori}
\medskip

When one moves branes in all three two-torus directions, there are
no new images and thus the symmetry breaking pattern is the same as that in Eq.
(\ref{Utwo}). In spite of having the same symmetry breaking pattern,
 the geometric interpretation of this configuration is  of course different
from that of the D-brane splitting in two two-tori
 since now the splitting takes place in all three toroidal directions. On the
 dual field theory side, we shall see that in this case the deformation in all
 three toroidal directions corresponds to giving VEV's to three superfields-moduli
in the adjoint representation.

In general, if we split $2(2k_i^1)$ D6-branes in any one of the three two-tori,
$2(4k_j^2)$ D6-branes in any two of the three two-tori, and
$2(4k_l^3)$ D6-branes in three two-tori, the gauge symmetry breaking
pattern is:
\begin{eqnarray}
U(N) &\to&   U(N-\sum_i  k^1_i -\sum_j 2 k^2_j -\sum_l 2 k_l^3 )
\times \prod_i U( k^1_i) \nonumber \\
&& \times \prod_j U( k^2_j) \times
 \prod_l  U(k_l^3) \label{General U breaking
pattern}~.~\,
\end{eqnarray}

In sum, in the three cases discussed the field theory corresponds
to giving VEVs to  one, two and three chiral superfields in the anti-symmetric
representation of $Sp(2N)$ or adjoint representation of
$U(N)$, respectively. A detailed discussion of this field
theoretical interpretation via
 Higgsing will be given   in the next Section.

\section{Higgsings in Field Theory}

\subsection{Higgsings for D-Branes Parallel to the O6-Planes}

The D6-brane geometric operations discussed in the previous
Section are in one-to-one correspondence with the  field theory
Higgs mechanism, by giving VEVs to three chiral superfields-moduli
living on D6-branes.
[These three adjoint (or anti-symmetric)
 chiral super-multiplets for the $U(N)$ (or $Sp(2N)$) group are generated from the
decomposition of the original adjoint vector representation of
D9-branes, which is caused by the dimensional reduction of
D9-branes down to D6-branes due to T-duality in three directions.]
For definiteness, we shall  focus on the relatively complicated
gauge symmetry breaking pattern
\begin{eqnarray}
Sp(4k) &\rightarrow &Sp(2k)\times Sp(2k) \nonumber\\
        & \rightarrow &Sp(2k) \nonumber\\
        & \rightarrow &U(k)~,~\,
\end{eqnarray}
given by Eq. (\ref{USp breaking pattern3}). This pattern happens
when one moves D6-branes off the O6-planes in all two-tori
directions, which requires all three anti-symmetric chiral
supermultiplets to obtain VEVs, each one being responsible for one
breaking step. The
analysis will be very similar for the other gauge symmetry breaking patterns
in Eqs. (\ref{USp breaking pattern1}) and (\ref{USp breaking pattern2})
where one splits the D6-branes off the O6-planes
in one and two two-tori, respectively.

To start with, let us review some simple properties of $Sp$
groups. $Sp(2N)$ is defined as the group of pseudo orthogonal
transformations preserving the anti-symmetric inner product
\begin{eqnarray}
\eta_\alpha \Sigma_{\alpha\beta} \xi_\beta ~,~
\end{eqnarray}
where $\Sigma=\sigma_2\bigotimes I_{N\times N}$.
The generators must satisfy the algebraic condition
\begin{eqnarray}
T^T\Sigma+\Sigma T =0 ~.~
\end{eqnarray}
Their
representations then can be expanded in the direct product form
\begin{eqnarray}
T&=&I_{2\times 2} \bigotimes T_a + \vec{\sigma}
\bigotimes\vec{T_s}\nonumber\\
&&=\left(\matrix{{A_N} &{S_N^1} \cr
               {S_N^2} &{-A_N^T}} \right)~,~\label{generator expansion}
\end{eqnarray}
where $A_N=T_a+T_s^3$, $S_N^1=T_s^1-T_s^2$ and $S_N^2=T_s^1+T_s^2$
are $N \times N$ matrices
with $T_a=-T_a^T$ and $T_s=T_s^T$. For the special case $T_S^1=T_S^2=0$, the
representations are reduced to
\begin{eqnarray}
T=\left(\matrix{{A_N} &{0} \cr
               {0} &{-A_N^T}} \right)~,~
\label{U group}
\end{eqnarray}
leaving the $Sp(2N)$ subgroup $U(N)$. For convenience we have
taken a real
convention for the Pauli matrices,\footnote{Hermitian generators can be
obtained by taking appropriate linear combinations.}
%
\begin{eqnarray}
\sigma_1=\left(\matrix{{0} &{1} \cr
               {1} &{0}} \right), &
\sigma_2=\left(\matrix{{0} &{-1} \cr
               {1} &{0}} \right), &
\sigma_3=\left(\matrix{{1} &{0} \cr
               {0} &{-1}} \right).
\end{eqnarray}

It is convenient to employ a tensor language in which the indices
are raised by $\Sigma^T$ and lowered by $\Sigma$.
The invariant pseudo inner product then can be rewritten as
\begin{eqnarray}
\eta^\alpha\xi_\alpha~,~
\end{eqnarray}
which obeys the transformation laws
\begin{eqnarray}
\xi_\alpha \rightarrow \xi_\alpha+(T\xi)_\alpha ~,~\\
\eta^\alpha \rightarrow \eta^\alpha-(\eta T)^\alpha ~.~\,
\end{eqnarray}
There are three $N=1$ anti-symmetric chiral
supermultiplets for the $Sp$ gauge factor on
 filler branes. Raising one index of these
representations, they will obey the transformation law
\begin{eqnarray} \Phi_\alpha^\beta \rightarrow
\Phi_\alpha^\beta+(T\Phi)_\alpha^\beta-(\Phi T)_\alpha^\beta~,~
\end{eqnarray}
and can be expanded as
\begin{eqnarray}
\Phi_\alpha^\beta=I \bigotimes \Phi_s +\vec{\sigma}
\bigotimes\vec{\Phi}_a~,~\label{Phi exansion}
\end{eqnarray}
where $\Phi_s^T=\Phi_s$ and $\vec{\Phi}_a^T=-\vec{\Phi}_a$.
$\Phi_\alpha^\beta$ and the $Sp$ generators $T$ have different
expansion forms since they belong to different representations. By giving
$\Phi^a$, $\Phi^b$ and $\Phi^c$ VEVs
\begin{eqnarray}
\langle\Phi^a\rangle= I \bigotimes v^a \left(\matrix{{I_k} &{0}
\cr
               {0} &{-I_k}} \right)~,~\\
\langle\Phi^b\rangle= I \bigotimes v^b \left(\matrix{{0} &{I_k}
\cr
               {I_k} &{0}} \right)~,~\\
\langle\Phi^c\rangle= \sigma_3 \bigotimes v^c \left(\matrix{{0}
&{I_k} \cr
               {-I_k} &{0}} \right)~,~
\end{eqnarray}
the original $Sp(4k)$ symmetry will be broken down to $U(k)$ step
by step.  If we diagonalize the matrices
 $\langle\Phi^a\rangle$,
$\langle\Phi^b\rangle$ and $\langle\Phi^c\rangle$, we obtain two
 eigenvalues with the same magnitude but opposite
sign for each one, which indicates that the
D6-branes are split on each two-torus.

To discuss the gauge symmetry breaking,
let us explicitly write the direct product expansions for the
$Sp(4k)$ adjoint representation
\begin{eqnarray}
T=I \bigotimes \left(\matrix{{(T_a)_{11}} &{(T_a)_{12}} \cr
               {-(T_a)_{12}^T} &{(T_a)_{22}}} \right)
               +  \vec{\sigma}\bigotimes\left(\matrix{{(\vec{T_s})_{11}} &{(\vec{T_s})_{12}} \cr
               {(\vec{T_s})_{12}^T} &{(\vec{T_s})_{22}}}
               \right)~,~
\end{eqnarray}
where the diagonal entries of $T_a$ and $T_s$ are $k \times k$
matrices which satisfy $(T_a)_{ii}=-(T_a)_{ii}^T$,
$(\vec{T_s})_{ii}=(\vec{T_s})_{ii}^T$ and the off-diagonal entries
$(T_a)_{12}$, $(\vec{T_s})_{12}$ are arbitrary. Once $\Phi^a$,
$\Phi^b$ or $\Phi^c$ obtains VEVs, only those symmetries commuting
with $\langle\Phi^a\rangle$, $\langle\Phi^b\rangle$ or
$\langle\Phi^c\rangle$ survive.
 Given
\begin{eqnarray}
(X_1\bigotimes Y_1)(X_2\bigotimes Y_2)=(X_1X_2)\bigotimes
(Y_1Y_2)\\
Tr(X \bigotimes Y)=(TrX)(TrY)~,~
\end{eqnarray}
one obtains
\begin{eqnarray}
[T,\langle\Phi^a\rangle]&=& I\bigotimes 2v^a\left(\matrix{{0}
               &{-(T_a)_{12}} \cr
               {-(T_a)_{12}^T} &{0}} \right) \nonumber \\
               &&+  \vec{\sigma} \bigotimes 2v^a\left(\matrix{{0} &{-(\vec{T_s})_{12}} \cr
               {-(\vec{T_s})_{12}^T} &{0}} \right)~,~
\end{eqnarray}
for a given $T$.  
The surviving symmetries
satisfy
\begin{eqnarray}
(T_a)_{12}=(\vec{T_s})_{12}=0~,~
\end{eqnarray}
yielding $Sp(2k)\times Sp(2k)$.

Subsequently, a non-zero $\langle\Phi^b\rangle$ leads to
\begin{eqnarray}
[T,\langle\Phi^b\rangle]&=& I\bigotimes v^b\left(\matrix{{0}
               &{(T_a)_{11}-(T_a)_{22}} \cr
               {(T_a)_{22}-(T_a)_{11}} &{0}} \right) \nonumber \\
               &&+  \vec{\sigma}\bigotimes v^b\left(\matrix{{0} &{(\vec{T_s})_{11}-(\vec{T_s})_{22}} \cr
               {(\vec{T_s})_{22}-(\vec{T_s})_{11}} &{0}} \right)~.~\,
\end{eqnarray}
The commutator is invariant if and only if
\begin{eqnarray}
(T_a)_{11}=(T_a)_{22},& (\vec{T_s})_{11}=(\vec{T_s})_{22}~,~
\end{eqnarray}
which identifies the two surviving gauge symmetries, leaving
$Sp(2k)$.

Finally, if $\Phi^c$ also obtains a VEV, the new commutator with
the $Sp(2k)$ generators will be
\begin{eqnarray}
[T,\langle\Phi^c\rangle]&=&v^c [\vec{\sigma},
\sigma_3]\bigotimes\left(\matrix{{0} &{(\vec{T_s})_{11}} \cr
               {-(\vec{T_s})_{11}} &{0}} \right)~.~\,
\end{eqnarray}
It vanishes if and only if
\begin{eqnarray}
(T_s^1)_{11}=(T_s^2)_{11}=0~,~\,
\end{eqnarray}
leaving the surviving symmetries
\begin{eqnarray}
T=I \bigotimes\left(\matrix{{(T_a)_{11}} &{0} \cr
               {0} &{(T_a)_{11}}} \right)
               + \sigma_3 \bigotimes \left(\matrix{(T_s^3)_{11} &{0} \cr
               {0} &{(T_s^3)_{11}}} \right)~.~\,
\end{eqnarray}
Referring to Eq. (\ref{U group}), we see that this is the $U(k)$ group.

Next, let us check D- and F-flatness for these breaking patterns.
D-flatness can be directly seen from
\begin{eqnarray}
Tr[\langle\Phi^i\rangle^{\dag} T\langle\Phi^i\rangle]=0~,~
\end{eqnarray}
where $i$ runs over $a$, $b$ and $c$. F-flatness is easy to check,
too. The superpotential for $\Phi^i$'s is
\begin{eqnarray}
W={\rm Tr}(\Phi^a\Phi^b\Phi^c+\Phi^a\Phi^c\Phi^b)
\end{eqnarray}
instead of the $N=4$ commutator structure. From Eq. (\ref{Phi
exansion}) one has
\begin{eqnarray}
{\rm Tr}(\Phi^a\Phi^b\Phi^c)={\rm Tr}(\Phi^a\Phi^c\Phi^b)~.~\,
\end{eqnarray}
For example,
\begin{eqnarray}
{\rm Tr}[\Phi^a\langle\Phi^b\rangle\langle\Phi^c\rangle]=2v^bv^c
{\rm Tr}[\Phi_a^3\left(\matrix{{-I_k}
&{0} \cr
               {0} &{I_k}} \right)]=0~,~
\end{eqnarray}
where $\Phi_a^3$ is the third anti-symmetric component matrix of
$\vec{\Phi}^a$. Similarly, ${\rm Tr}[\langle
\Phi^a\rangle\Phi^b\langle\Phi^c\rangle]=
{\rm Tr}[\langle\Phi^a\rangle\langle\Phi^b\rangle\Phi^c]=
{\rm Tr}[\langle\Phi^a\rangle\langle\Phi^b\rangle\langle\Phi^c\rangle]=0$,
so F-flatness is also preserved.

This result is consistent with the  symmetry breaking pattern due
to the brane splitting in the string theory  constructions. The F- and
D-flatness preserving VEVs for one, two and three chiral
superfields in the anti-symmetric representation of the original
$Sp$ group  is in one-to-one correspondence with the symmetry
breaking pattern due to the splitting of D6-branes, parallel with
the O6-planes  in one-, two- and three- two-torus directions,
respectively. In other words, the three chiral superfields  are the
moduli associated with the parallel motion of D6-branes wrapping
(non-rigid) cycles, parallel with the O6-planes. Note that the D- and
F-flatness are not automatically guaranteed for Higgsing if two or three
 chiral superfields in the anti-symmetric representation
of $Sp$ branes (or adjoint representation for $U$ branes) obtain VEVs.
However, the
 D- or F-flatness violating Higgsing does not typically correspond to a
consistent deformation of D6-brane configurations.

\subsection{Higgsings for D-Branes Not Parallel to the O6-Planes}

For the sake of completeness, in this subsection
  we shall also discuss the Higgsing for the D6-branes not parallel to the
  O6-planes. The string theory   aspects of
  symmetry breaking  via such D6-brane splitting have been discussed in Section
  2.2. Here we  give the dual field theory  description
  of the  symmetry breaking
via the supersymmetry preserving  Higgs mechanism due to the three
chiral superfields in the adjoint representation of $U(N)$.

Recall, the $U(N)$ symmetry breaking chain  due to brane-splitting is:
\begin{eqnarray}
U(N) &\rightarrow & U(N-k)\times U(k) \nonumber\\
        & \rightarrow & U(N-2k)\times U(k)  \nonumber\\
        & \rightarrow & U(N-2k)\times U(k) ~.~\,
\end{eqnarray}

In the canonical generator basis for the $U(N)$ group, we choose the
VEVs of $\Phi^a$, $\Phi^b$ and $\Phi^c$ as
\begin{eqnarray}
\langle \Phi^a \rangle ~=~ \left(\matrix{ - v_a I_{k \times k} & 0_{k \times (N-k)} \cr
0_{ (N-k) \times k} &  v_a I_{ (N-k) \times (N-k) } } \right)~,~\,
\end{eqnarray}
\begin{eqnarray}
\langle \Phi^b \rangle = \left(\matrix{ 0_{k \times k} & v_b I_{k \times k}
 & 0_{k \times (N-2k)} \cr
v_b I_{k \times k} & 0_{k \times k} & 0_{k \times (N-2k)} \cr
0_{(N-2k) \times k} & 0_{(N-2k) \times k}
& 0_{(N-2k) \times (N-2k)} } \right)~,~\,
\end{eqnarray}
\begin{eqnarray}
\langle \Phi^c \rangle = \left(\matrix{ 0_{k \times k} & -i v_c I_{k \times k}
 & 0_{k \times (N-2k)} \cr
i v_c I_{k \times k} & 0_{k \times k} & 0_{k \times (N-2k)} \cr
0_{(N-2k) \times k} & 0_{(N-2k) \times k}
& 0_{(N-2k) \times (N-2k)} } \right)~,~\,
\end{eqnarray}
where $I_{n\times n}$ is the $n\times n$ identity matrix,
and $0_{n\times m}$ is the $n\times m$ matrix in which
all the entries are zero.

It can easily be shown that $\Phi^a$ breaks the
$U(N)$ gauge symmetry  down to $U(N-k)\times U(k)$, and
 $\Phi^b$ breaks $U(N-k)\times U(k)$ to $U(N-2k)\times U(k)$.
However, $\Phi^c$ does not lead to a further breaking of the gauge symmetry,
similar to the gauge symmetry breaking
via brane splitting.

D-flatness can be obtained from
\begin{eqnarray}
D^a \equiv {\rm Tr} \left(\langle\Phi^{i}\rangle^{\dag} [T^a, \langle\Phi^i\rangle]\right)=0~,~
\end{eqnarray}
where $i$ runs over $a$, $b$ and $c$, and $T^a$ is the
generator of $U(N)$.

The superpotential for $\Phi^i$'s can be written as (see e.g.,
\cite{CSU2}):
\begin{eqnarray}
W={\rm Tr} \left(\Phi^a\Phi^b\Phi^c+\Phi^b\Phi^a\Phi^c\right) ~.~\,
\end{eqnarray}
The superpotential can easily be seen to vanish when any two
(or all three) of $\Phi^a$, $\Phi^b$ and $\Phi^c$
 obtain VEVs,
demonstrating
F-flatness.

\section{Standard-like Models and Brane-Splitting}

One natural way to obtain the
three-family $N=1$ supersymmetric standard-like models via D6-brane-splitting
was addressed in \cite{CLL}.  One started with the
Pati-Salam gauge symmetry  $U(4)\times U(2)_L\times U(2)_R$ in the
observable sector. The left-right symmetric model
$SU(3)\times SU(2)_L \times SU(2)_R \times U(1)_{B-L}$   was then obtained
 by splitting $U(4)$ branes, and the gauge symmetry $SU(3)_C\times
SU(2)_L\times U(1)_{B-L} \times U(1)_R$  was obtained
by further splitting
$U(2)_R$ branes. Finally, the Standard Model
gauge symmetry  emerged  by  giving VEV's to the
 massless open string states in an $N=2$
subsector, {\it i.e.}, the field theory analog of brane recombination.

In this Section, we shall  further explore the construction of
the Standard-like models, by employing  the D6-brane splitting mechanism for
the $\IZ_2\times \IZ_2$ orientifold models.
Here, the electroweak symmetry will emerge
  primarily from the splitting of D6-branes
 parallel with the O6-planes.
 Within the supersymmetric
 constructions with intersecting D6-branes, this possibility has not been
 explored, yet.
 Such Standard-like  models  shall be generated for two
Pati-Salam-like models. For the first example we shall employ the D-brane
splitting as discussed in Sections 3 and 4; this procedure of course preserves
 supersymmetry and  corresponds to the consistent string construction.
  For the second model we will present  the
Higgs mechanism that breaks it down to the three family Standard Model,
but
unfortunately it  does not preserve the D- and F-flatness condition.  This Higgsing
procedure  therefore
does not have an interpretation in terms of a  consistent  D-brane splitting
construction.
 The main difference in comparison with the models discussed in \cite{CLL}
 is that  now the  starting  electroweak symmetry is  based on $Sp$ groups, {\it i.e.},
 $U(4)\times Sp(8\ {\rm or}\ 6)_L \times Sp(8\ {\rm or}\  6)_R$, and not on $U$
 groups, {\it i.e.},
$U(4)\times U(2)_L\times U(2)_R$, as in \cite{CLL}. We shall allow
for the brane splitting  for all three brane  stacks, leading to
$SU(3)\times Sp(2)_L \times Sp(2)_R \times U(1)_{B-L}$, again. In
particular, for the four-family supersymmetric model, its
observable sector and the hidden sector do not spatially
intersect, therefore yielding no massless exotic particles.

The third model presented in Section 4.3 is the  three-family
Standard-like model  based directly on the $Sp(2)_L\times Sp(2)_R$
brane constrcution. It is related to the local construction in
\cite{cim5,cim6}. Our construction is  a consistent supersymmetric
one, but it suffers from the Standard Model chiral exotics.

\subsection{Model I: Four Family Standard-like Model}
\begin{table}
[htb] \footnotesize
\renewcommand{\arraystretch}{1.0}
\caption{D6-brane configurations and intersection numbers for the
four-family Standard-like model. In the table,
$\chi_i$ is the complex modulus for the $i-$th torus, and
$\beta_i^g$ is the beta function for the $i-$th $Sp$ group
 from the $i-$th stack of branes.} \label{four-family model}
\begin{center}
\begin{tabular}{|c||c|c||c|c|c|c|c|c|c|c|}
\hline
    \rm{I} & \multicolumn{10}{c|}{$[U(4)_C\times Sp(8)_L\times
    Sp(8)_R]_{observed} \times [U(4)\times Sp(8)\times Sp(8)]_{hidden}$}\\
\hline \hline \rm{stack} & $N$ & $(n^1,l^1)\times (n^2,l^2)\times
(n^3,l^3)$ & $n_{\Ysymm}$& $n_{\Yasymm}$ & $b$  & $c$ & $d$ & $d'$ & 1 & 2  \\
\hline
\hline
    $a$&  8& $(1,0)\times (1,1)\times (1,-1)$ & 0 & 0  &  1 & -1 & 0 & 0 & 0 & 0 \\
    $b$&  8& $(0,1)\times (1,0)\times (0,-1)$ & 0 & 0  &  - & 0 & 0 & 0 & 0 & 0 \\
    $c$&  8& $(0,1)\times (0,-1)\times (1,0)$ & 0 & 0  &  - & - & 0 & 0 & 0 & 0 \\
\hline
\hline
    $d$&   8& $(0,1)\times (1,-1)\times (1,-1)$ & 0 & 0  &  - & - & - & 0 & -1 & 1 \\
\hline
    1&   8& $(1,0)\times (1,0)\times (1,0)$  & \multicolumn{8}{c|} {$\chi_2=\chi_3=1$}\\
    2&   8& $(1,0)\times (0,-1)\times (0,1)$ & \multicolumn{8}{c|} {$\beta^g_1= \beta^g_2=-4$}\\
\hline
\end{tabular}
\end{center}
\end{table}

Model I is a four-family model, with its D6-brane configurations
and intersection numbers listed in Table \ref{four-family
model}. This model contains $U(4)_C\times Sp(8)_L\times
Sp(8)_R$ and $U(4)\times Sp(8)\times Sp(8)$ gauge structures in
its observable and hidden sectors, respectively. [For technical details and
consistency conditions for  RR-tadpole free supersymmetric constructions see
\cite{CSU2,CPS,CLL}.]
 From the  wrapping
numbers of the stacks of D6-branes, it is not hard
 to check that no intersection happens
between these two sectors and thus the chiral  Standard Model exotic particles,
which had been a generic problem for supersymmetric constructions with
intersecting D6-branes,  are
absent here.

Let us focus on the observable sector first. As discussed in the previous
Sections the $Sp(4k)$
group can be broken down to $U(k)$ symmetry by moving D6-branes
off O6-planes in all two-tori directions. This indicates that the
$Sp(8)_L\times Sp(8)_R$ gauge factors in this model can be
diagonally broken down to $U(2)_L \times U(2)_R$.

In this case both
$U(1)$ factors are non-anomalous since they arise from the non-Abelian, {\it i.e.},
$Sp$, symmetry. (All non-Abelian gauge symmetries in these string constructions are
non-anomalous \cite{CSU2}.)
Meanwhile, the original chiral supermultiplets , $i.e.$,
$(4,8,1;1,1,1)$ and $(4,1,8;1,1,1)$ are decomposed into four
identical ones, $i.e.$, $(4,2,1;1,1,1)$ and $(4,1,2;1,1,1)$,
respectively.
Note that the four families in the effective theory
have been obtained from a single family in the original construction.
The fact that the number of families can change for this type of
 Higgsing  has the
 string origin in the original configuration on
the top of the O6-plane, {\it i.e.}, the {\it
singular}  configuration, fixed by  the orientifold action.
The subsequent splitting of  branes
away from this  singularity in turn allows for the change
 in the number of chiral families.
 Furthermore,  this specific Higgsing, as discussed in  the previous
Section,  preserves   D- and
F-flatness and thus the breaking can take place at
  the string scale. Namely, in the dual  string theory, the
   brane-splitting can take place at any separation of branes, limited only by
   the size of the internal space.

After the brane-splitting, as discussed above,
 the resulting  ``observable  sector''   is therefore
  a four-family Pati-Salam model $U(4)\times
U(2)_L\times U(2)_R$.
As for the ``hidden sector'', all the $Sp$ gauge factors have negative $\beta$ functions
and therefore allow for  confinement  at some intermediate
scale and  gaugino condensation there. This mechanism
 would in turn generate a
non-perturbative effective superpotential compactification moduli, which could
allow for the ground state solution with stabilized moduli.
 For more details, see
\cite{CLW}.

\subsection{Model II: Non-Supersymmetric Three Family Standard Model}
\begin{table}
[htb] \footnotesize
\renewcommand{\arraystretch}{1.0}
\caption{D6-brane configurations and intersection numbers for the
three-family Standard-like model. In the table, $\chi_i$ is the
complex modulus for the $i-$th torus, and $\beta_i^g$ is the beta
function for the $i-$th $Sp$ group from the $i-$th stack of
branes. The third torus is tilted, so $l^3=n^3+2m^3$\cite{CPS}. }
\label{three-family model}
\begin{center}
\begin{tabular}{|c||c|c||c|c|c|c|c|c|c|c|}
\hline
    \rm{II} & \multicolumn{10}{c|}{$[U(4)_C\times Sp(6)_L\times
    Sp(6)_R]_{observed} \times [U(2)\times Sp(4)\times Sp(4)]_{hidden}$}\\
\hline \hline \rm{stack} & $N$ & $(n^1,l^1)\times (n^2,l^2)\times
(n^3,l^3)$ & $n_{\Ysymm}$& $n_{\Yasymm}$ & $b$  & $c$ & $d$ & $d'$ & 1 & 2  \\
\hline \hline
    $a$&  8& $(1,0)\times (1,1)\times (1,-1)$ & 0 & 0  &  1 & -1& 0 & 0  & 0 &0 \\
    $b$&  6& $(0,1)\times (1,0)\times (0,-2)$ & 0 & 0  &  - & 0 & 0  & 0 & 0 & 0 \\
    $c$&  6& $(0,1)\times (0,-1)\times (2,0)$ & 0 & 0  &  - & - & 0  & 0 & 0 & 0 \\
\hline \hline
    $d$&4&$(0,1)\times (1,-1)\times (1,-1)$ &\multicolumn{8}{c|} {$\chi_2=\frac{1}{2}\chi_3=1$}\\
    1&   4& $(1,0)\times (1,0)\times (2,0)$  &\multicolumn{8}{c|}{$\beta^g_1=-5$, $\beta^g_2=-5$} \\
    2&   4& $(1,0)\times (0,-1)\times (0,2)$ & \multicolumn{8}{c|}{}\\
\hline
\end{tabular}
\end{center}
\end{table}

Model II is a three-family model, with its D6-brane configurations
and intersection numbers given in Table \ref{three-family model}.
This model contains $U(4)_C\times Sp(6)_L\times Sp(6)_R$  and
$U(2) \times Sp(4)\times Sp(4)$ gauge structures in its observable
and hidden sectors, respectively. This model also shares the
advantage of Model I, $i.e.$, there are no massless Standard Model chiral
exotics since the branes in the observable and hidden sectors
do not intersect at a point in the internal space.
[First version had chiral exotics and suffered  from the
global anomaly  \cite{WITT}, due to the odd number of fundamental
representations under $Sp(2N)$ symmetries.]

However, in this model the Higgsing  down to the Standard Model symmetry
with three-families  breaks supersymmetry since  such a  symmetry breaking
pattern does not  preserve the  F- and D-flatness conditions.
Alternatively, there is no pallel D-brane splitting  that would
result  in  such a symmmetry breaking pattern. Nevertheless let us
consider one  typical breaking pattern:
\begin{eqnarray}
Sp(6) &\rightarrow &Sp(2)\times Sp(2)\times Sp(2) \nonumber\\
        & \rightarrow &Sp(2) \times Sp(2) \nonumber\\
        & \rightarrow &Sp(2)~.~\,  \label{tevscale}
\end{eqnarray}
This can be realized by choosing VEVs for three
anti-symmetric chiral supermultiplets on   branes $\Phi_a$,
$\Phi_b$ and $\Phi_c$:
\begin{eqnarray}
\langle\Phi^a\rangle= I_{2\times 2} \bigotimes v^a
\left(\matrix{{1} &{0}&{0}\cr {0} &{-1}&{0} \cr
{0}&{0}&{0}} \right)~,~\\
\langle\Phi^b\rangle= I_{2\times 2} \bigotimes v^b
\left(\matrix{{0} &{1}&{0}\cr {1} &{0}&{0} \cr
{0}&{0}&{0}} \right)~,~\\
\langle\Phi^c\rangle= I_{2\times 2} \bigotimes v^c
\left(\matrix{{0} &{0}&{0}\cr {0} &{0}&{1} \cr {0}&{1}&{0}}
\right)~,~
\end{eqnarray}
where $\langle\Phi_a\rangle$ breaks $Sp(6)$ into $Sp(2)^3$;
$\langle\Phi_b\rangle$ identifies the first two $Sp(2)$ factors
and $\langle\Phi_c\rangle$ identifies the last two. We can also
choose:
\begin{eqnarray}
\langle\Phi^b\rangle= I_{2\times 2} \bigotimes v^b
\left(\matrix{{0} &{1}&{0}\cr {1} &{0}&{1} \cr {0}&{1}&{0}}
\right)
\end{eqnarray}
to break $Sp(2)^3$ directly down to the diagonal $Sp(2)$. This
Higgsing  yields the three-family  Standard Model, where the three
families emerge from a single family  in the original string
construction. However, let us emphasize again that this
proceduere, since it breaks supersymmetry, does not correspond to
a consistent D-brane splitting mechanism. F-flatness cannot be
preserved even though D-flatness is. By the way, if only $\Phi^a$
and $\Phi^b$ (or $\Phi^a$ and $\Phi^c$) obtain VEVs, the $Sp(6)$
gauge symmetry is broken down to $Sp(2)\times Sp(2)$, and the D-
and F-flatness can be preserved, {\it i.e.}, this is the
supersymmetry preserving Higgs mechanism.

The model could still be  viable if $v^a$, $v^b$, and
$v^c$ are all at  the TeV scale where supersymmetry is expected to be broken
(choosing one or two at a higher scale would introduce a new hierarchy problem).
However, this introduces a new difficulty: since the three families
emerge from the breaking pattern in Eq. (\ref{tevscale}), the
TeV scale gauge bosons of the broken $Sp(6)$ can mediate
potentially dangerous flavor changing neutral (and charged) current transitions between the
three families. A detailed investigation would require a full
knowledge of the fermion masses and mixings in the model, and is beyond
the scope of this paper.
However, the decays $K_L \rightarrow \mu^{\pm} e^{\mp}$ are especially
dangerous since they can occur even in the absence of family mixing. The experimental limit
$B < 5 \times 10^{-12}$ on the branching ratio~\cite{pdg} suggests a lower
limit of $\sim (50-100)$ TeV on the masses of these gauge bosons, which is rather high
compared to the supersymmetry breaking scale that is usually assumed.

However, even higher scales may be allowable.
From the  study of the landscape  of string theory vacua, {\it i.e.}, the statistical
study of string theory vacua, it was argued recently
that the most numerous ``acceptable vacua'' do not have the low
energy supersymmetry~\cite{Susskind:2004uv, Douglas:2004qg},
 {\it i.e.}, the low energy supersymmetry is
not natural, and a phenomenological model has been proposed in which
supersymmetry is broken at an intermediate scale \cite{Arkani-Hamed:2004fb}.
 If this were the case this model could  be very interesting.
The anomalies from $U(1)$ in $U(4)$ are cancelled by the
Green-Schwarz mechanism, and the gauge field of $U(1)$ obtains
mass via the linear $B\wedge F$ couplings, and thus in the
observable sector  the gauge symmetry  (with massless gauge
bosons) is $SU(4)_C\times Sp(6)_L\times Sp(6)_R$. In addition,
$SU(4)_C$ gauge symmetry can be broken down to the $SU(3)_C\times
U(1)_{B-L}$ via brane splitting. And the $Sp(6)_L$ and $Sp(6)_R$
can be broken down to $SU(2)_L$ and $SU(2)_R$ ($SU(2)\equiv
Sp(2)$) via the supersymmetry breaking Higgs mechanism, as
discussed above, at the intermediate scale, say, $10^{12}$ GeV.
Moreover, the $SU(2)_R\times U(1)_{B-L}$ gauge symmetry can be
broken down to $U(1)_Y$ by the VEVs of the scalar components in
the right-handed neutrino superfields, which could also  take
place at the  intermediate scale, say,  $10^{12}$ GeV. According
to the string landscape arugment such a solution may in turn
correspond to a viable string vacuum solution.


\subsection{Model III based on $Sp(2)_L\times Sp(2)_R$}
Recently, the locally supersymmetric three-family Standard Model
with only one pair of Higgs doublets was presented in
\cite{cim5,cim6}. The construction is based on the toroidal Type
IIA orientifold with intersecting D6-branes. The electroweak part
of the Standard Model is based on $Sp(2)_L\times Sp(2)_R$ gauge
symmetry and arises from D6-branes parallel with the
O6-orientifold planes.
 Unfortunately, the D6-brane
configurations of the Standard Model  do not cancel the RR tadpoles, which
could be cancelled by adding anti-D6-branes. However, in this case
the full  model is not supersymmetric anymore \cite{cim5,cim6}.

\begin{table}
[htb] \footnotesize
\renewcommand{\arraystretch}{1.0}
\caption{D6-brane configurations and intersection numbers for the
supersymmetric model whose observable sector is similar to the
locally supersymmetric Standard Model in \cite{cim5,cim6}. In the
table, $\chi_i$ is the complex modulus for the $i-$th torus, and
$\beta_i^g$ is the beta function for the $i-$th $Sp$ group
 from the $i-$th stack of branes. The third two-torus is tilted, so
$l^3=n^3+2m^3$\cite{CPS}} \label{lSSM}
\begin{center}
\begin{tabular}{|c||c|c||c|c|c|c|c|c|c|}
\hline
    \rm{III} & \multicolumn{9}{c|}{$[U(4)_C\times SU(2)_L\times
    SU(2)_R]_{observable} \times [U(2)\times Sp(8)]_{hidden}$}\\
\hline \hline \rm{stack} & $N$ & $(n^1,l^1)\times (n^2,l^2)\times
(n^3,l^3)$ & $n_{\Ysymm}$& $n_{\Yasymm}$ & $b$  & $c$ & $d$ & $d'$  & 2  \\
\hline
\hline
    $a$&  8& $(1,0)\times (1,3)\times (1,-3)$ & 0 & 0  &  3 & -3&0  & 0 & 0  \\
    $b$&  2& $(0,1)\times (1,0)\times (0,-2)$ & 0 & 0  &  - & 0 &-6 & 6 & 0  \\
    $c$&  2& $(0,1)\times (0,-1)\times (2,0)$ & 0 & 0  &  - & - & -6& 6 & 0  \\
\hline
\hline
    $d$&  4 & $(2,-1)\times (1,3)\times (1,3)$ & \multicolumn{7}{c|} {$\chi_1=24\chi_3/(4-9\chi_3^2)$}\\
    2  &$8$& $(1,0)\times (0,-1)\times (0,2)$ & \multicolumn{7}{c|} {$\chi_2=\frac{1}{2}\chi_3$, $\beta^g_2=-5$}\\
\hline
\end{tabular}
\end{center}
\end{table}

Interestingly, this specific Standard Model sector can be
implemented into the consistent supersymmetric constructions based
on the
 Type IIA $T^6/(\IZ_2\times \IZ_2)$ orientifold.
The models have  an  additional  gauge
 sector and  unfortunately possess  the Standard Model chiral exotics that appear
 at the intersections of the Standard Model D6-branes with those of the
 additional gauge sector.

Here we present one such model, with the third two-torus tilted.
The D6-brane configurations and intersection numbers are given  in
Table \ref{lSSM}. [The model  presented in the first version had a $U(1)$
factor in the hidden  sector  which introduced  odd-number of
fundamental representations  of the $Sp(2N)$
symmetries, and thus  the model possessed the  global anomaly
\cite{WITT}; the conditions for  the absence of such global anomalies can
be  derived  from K-theory \cite{UR} and were given  for
Type  IIA $T^6/(\IZ_2\times \IZ_2)$ orientifolds in \cite{MS}.]
We  assume that the $b$ and $c$ stacks of
D6-branes coincide with each other on the first torus. The
Standard Model D6-branes have the same wrapping numbers as those
of \cite{cim5,cim6}. In the observable sector  the gauge symmetry
is $U(4)\times Sp(2)_L\times Sp(2)_R$, where the anomalies from
$U(1)$ in $U(4)$ are cancelled by the Green-Schwarz mechanism, and
the gauge field of $U(1)$ obtains mass via the linear $B\wedge F$
couplings. Using the equivalence $Sp(2)\equiv SU(2)$, the
 gauge  symmetry  (with massless gauge bosons) is   then $SU(4)_C\times SU(2)_L\times SU(2)_R$,
{\it i.e.}, the Pati-Salam model.  We can  further break
the $SU(4)_C$ symmetry down to the $SU(3)_C\times U(1)_{B-L}$ via
the D6-brane splitting,  which leads to
 $SU(3)\times SU(2)_L\times SU(2)_R \times U(1)_{B-L}$  gauge symmetry at the
 string scale. Since in the Standard Model sector the D6-branes wrap the
 same three-cycles as those in \cite{cim5,cim6}, there are
 three families of the Standard Model fermions, and one pair of
Higgs doublets arising from the open strings which stretch between
the $b$ and $c$ stacks of D6-branes.
However, in order to cancel the RR tadpoles, we have to introduce
an additional  gauge sector, in particular, one $U(2)$ and one
$Sp$ groups. It turns out that all the Standard Model chiral
exotic particles are charged under this $U(2)$, {\it i.e.}, they
appear at the intersection of the Standard Model D6-branes and the
$U(2)$ D6-brane. The hope is that they may become massive
 after the $U(2)$ symmetries are broken. In this case  at  low energies this model
 would be  very attractive:   $SU(2)_R\times U(1)_{B-L}$ gauge symmetry can
be broken down to the $U(1)_Y$ by the VEVs of
 the scalar components in the right-handed
neutrino superfields at several TeV or at an intermediate scale
(if one accepts the landscape arguments~\cite{Susskind:2004uv,
Douglas:2004qg}). Therefore at the electroweak scale one would
only have the MSSM-like vacua.

This is only a representative model in this class and one can
construct variants of the above model by choosing different
wrapping numbers for branes in the gauge sector beyond the
Standard Model. However, these models typically possess
additional Standard Model exotics.

\section{Conclusions}

In this paper, we addressed in detail the string  and  field theory aspects of
parallel
D-brane splitting in Type IIA orientifolds.
This is a specific phenomenon due to
the fact that for toroidal orbifolds the cycles wrapped by D-branes are
generically  not
rigid and thus the continuous parallel splitting of branes  (consistent with
preservation of supersymmetry)  can take place. Non-rigid cycles are generic for
toroidal/orbifold compactifications, where the conformal field theory techniques can
be employed to obtain the spectrum and couplings for four-dimensional models
with potentially interesting particle physics properties. Especially,
the constructions with the
intersecting D6-branes on orientifolds of that type have interesting particle
physics implications.

 For the sake of concreteness we
analyse  the D6-brane splitting on ${\bf T}^6/(\IZ_2\times \IZ_2)$ orientifold
(with factorizable ${\bf T}^6={\bf T}^2\times {\bf T}^2\times{\bf T}^2$).
In  particular,  the symmetry breaking pattern
for D6-branes parallel with the O6-planes is nontrivial, since it involves a
combination of the orientifold and orbifold projections on Chan-Paton indices  for
open string states  as well as  putting branes in different
configurations, related to each other by the geometric actions of
the orientifold and orbifold group
elements. We analysed all the possible symmetry breaking patterns with D6-brane
splitting in one-, two- and three- two-tori, respectively. In a field theory we
demonstrate one-to-one correspondence with the supersymmetry preserving
 Higgs mechanism  by giving VEVs to one-,
two- and three- chiral superfields, respectively. These superfields
  in the anti-symmetric representation of the
original $Sp$ symmetry  are the moduli associated with the parallel splitting
of the D6-branes.

 This symmetry breaking mechanism within the intersecting D6-brane construction
 allows for the change in the number of chiral families, which appear at the
 intersections of the split-branes with another stack of branes.
 This change in the
 number of chiral superfields  is due to the fact that the original brane configuration
 on the top of the O6-plane was singular (fixed plane of the orientifold action) and
 thus the deformation from such a singular D-brane configuration allows for
 the change in the number of chiral  fields.

 For the sake of completeness we also address the symmetry breaking pattern for
 the branes not parallel with the orientifold planes,  both
 in the string theory and field theory. In this case the Higgsing is due to the
 three chiral superfields in the adjoint representation of the original $U$
 symmetry.

As an application of this  analysis we present three examples of new
supersymmetric Standard-like models where the end-point gauge symmetry and the
spectrum is a consequence of brane-splitting, both those parallel with the
orientifold planes and those that are not. In particular, we construct a
four-family  model with no Standard Model chiral exotics where the electroweak
symmetry is obtained from the $Sp(2N)_L\times Sp(2N)_R$  original brane
configuration.

We also address a three-family model, again with no massless SM
chiral exotics, in which the Standard Model is obtained by
non-supersymmetric Higgsing, which however does not have a
consistent string theory interpretation in terms of D-brane
deformations. Such a model may be viable if the Higgsing occurs at
a low or intermediate scale. However, the scale should be at least
$\sim(50-100)$ TeV to suppress unwanted flavor changing neutral
currents.

In addition, we present a  consistent supersymmetric
 three-family  Standard-like model  with one vector pair of Higgs doublets and
 the  $Sp(2)_L\times
Sp(2)_R$ electroweak symmetry arising in  the original string construction.
The wrapping numbers of the branes in  the  Standard Model
sector are the same as those in   \cite{cim5,cim6}, however, in our
construction we found explicit additional brane configurations that
cancel all the RR-tadpoles. On the other hand, these additional gauge
sectors  typically  introduce Standard  Model chiral exotics.

The study  presented in this paper porvides a  useful  tool for
further  constructions of intersecting  D6-brane models models
with potententially  realistic particle physics.

\section*{Acknowledgments}

We would like to thank A.~M.~Uranga for useful discussions and G.
Shiu for helpful comments. The research was supported in part by
the National Science Foundation under Grant No.~INT02-03585 (MC)
and PHY-0070928 (T. Li),  by the Department of Energy Grant
DOE-EY-76-02-3071 (MC, PL, T. Liu) and the Fay R. and Eugene L.
Langberg Chair  (MC).


\begin{thebibliography}{99}

%

\bibitem{AN}
C.~Angelantonj, I.~Antoniadis, E.~Dudas and A.~Sagnotti,
Phys.\ Lett.\ B {\bf 489}, 223 (2000), hep-th/0007090.

\bibitem{LU}
R.~Blumenhagen, L.~G\"orlich, B.~K\"ors and D.~L\"ust, JHEP {\bf
0010} (2000) 006,hep-th/0007024.

\bibitem{IB}
G.~Aldazabal, S.~Franco, L.~E.~Ib\'a\~nez, R.~Rabad\'an and
A.~M.~Uranga, JHEP {\bf 0102}, 047 (2001), hep-ph/0011132.


\bibitem{IB2}
G.~Aldazabal, S.~Franco, L.~E.~Ib\'a\~nez, R.~Rabad\'an and
A.~M.~Uranga, J.\ Math.\ Phys.\  {\bf 42}, 3103 (2001), hep-th/0011073.

\bibitem{LU2}
R.~Blumenhagen, B.~K\"ors and D.~L\"ust, JHEP {\bf 0102} (2001)
030,
hep-th/0012156.


\bibitem{CSU1}
M.~Cveti\v c, G.~Shiu and A.~M.~Uranga, Phys.\ Rev.\ Lett.\  {\bf
87}, 201801 (2001), hep-th/0107143.


\bibitem{CSU2}
M.~Cveti\v c, G.~Shiu and A.~M.~Uranga, Nucl.\ Phys.\ {\bf B615},
3 (2001), hep-th/0107166.


\bibitem{douglas}
M.~Berkooz, M.~R.~Douglas and R.~G.~Leigh, Nucl. Phys. B {\bf 480}
(1996) 265, hep-th/9606139.


\bibitem{imr}
L.~E.~Ib\'a\~nez, F.~Marchesano and R.~Rabad\'an, JHEP {\bf 0111},
002 (2001), hep-th/0105155.



\bibitem{bklo}
R.~Blumenhagen, B.~K\"ors and D.~L\"ust, T.~Ott, Nucl. Phys. {\bf
B616} (2001) 3,hep-th/0107138.



\bibitem{cim}
D.~Cremades, L.~E.~Ib\'a\~nez and F.~Marchesano, Nucl.\ Phys.\  {\bf
B643}, 93 (2002), hep-th/0205074.

\bibitem{CIM1}
D.~Cremades, L.~E.~Ib\'a\~nez and F.~Marchesano, JHEP 0207, 022 (2002),
hep-th/0203160.

\bibitem{bailin}
D.~Bailin, G.~V.~Kraniotis, and A.~Love, Phys.\ Lett.\ B {\bf
530}, 202 (2002); Phys.\ Lett.\ B {\bf 547}, 43 (2002); Phys.\
Lett.\ B {\bf 553}, 79 (2003); JHEP {\bf 0302}, 052 (2003).

\bibitem{JREDVN}
J.~R.~Ellis, P.~Kanti and D.~V.~Nanopoulos, Nucl.\ Phys.\ B {\bf
647}, 235 (2002).

\bibitem{kokorelis}  C.~Kokorelis,
JHEP {\bf 0209}, 029 (2002); JHEP {\bf 0208}, 036 (2002);
hep-th/0207234; JHEP {\bf 0211}, 027 (2002); hep-th/0210200.

\bibitem{LL}
T.~Li and T.~Liu, Phys.\ Lett.\ B {\bf 573}, 193 (2003),
hep-th/0304258.

\bibitem{CLS1}
 M.~Cveti\v c, P.~Langacker and G.~Shiu,
Phys.\ Rev.\ D {\bf 66}, 066004 (2002), hep-ph/0205252.

\bibitem{CLS2}
M.~Cveti\v c, P.~Langacker and G.~Shiu, Nucl.\ Phys.\ B {\bf 642},
139 (2002), hep-th/0206115.

\bibitem{CLW}
M.~Cveti\v c, P.~Langacker and J.~Wang, Phys.\ Rev.\ D {\bf 68},
046002 (2003).


\bibitem{CP2}
M.~Cveti\v c and I.~Papadimitriou, Phys.\ Rev.\ D {\bf 68}, 046001
(2003),hep-th/0303083.

\bibitem{CP} M. Cveti\v c and I. Papadimitriou,
Phys.\ Rev.\   {\bf D67}, 126006 (2003), hep-th/0303197.

\bibitem{CPS}
M. Cveti\v c, I. Papadimitriou and G. Shiu, Nucl.\ Phys.\  {\bf B659}, 193 (2003)
 hep-th/0212177.
\bibitem{Blum}
R.~Blumenhagen, L.~G\"orlich and T.~Ott,
JHEP {\bf 0301}, 021 (2003), hep-th/0211059.
\bibitem{Hon}
G.~Honecker, Nucl.\ Phys.\  {\bf B666}, 175 (2003), hep-th/0303015.
\bibitem{HO}
G.~Honecker and T.~Ott,
hep-th/0404055.


\bibitem{CLL}
M.~Cveti\v c, T.~Li and T.~Liu,
hep-th/0403061.

\bibitem{Blum2}
R.~Blumenhagen, JHEP {\bf0311}, 055 (2003),
hep-th/0310244.

\bibitem{Ilke}
I.~Brunner, K.~Hori, K.~Hosomichi and J.~Walcher,
hep-th/0401137.

\bibitem{BW}
R.~Blumenhagen and T.~Weigand, JHEP {\bf0402}, 041 (2004), hep-th/0401148.

\bibitem{schell}
T.~P.~T.~Dijkstra, L.~R.~Huiszoon and A.~N.~Schellekens,
hep-th/0403196.
\bibitem{LUS}
I.~Brunner, A.~Hanany, A.~Karch and D.~L\"ust,
Nucl.\ Phys.\ B {\bf 528}, 197 (1998), hep-th/9801017.

\bibitem{EM}
J.~Erdmenger, Z.~Guralnik, R.~Helling and I.~Kirsch,
branes,''
JHEP {\bf 0404}, 064 (2004), hep-th/0309043.

\bibitem{AS}
C.~Angelantonj and A.~Sagnotti,
Phys.\ Rept.\  {\bf 371}, 1 (2002)
[Erratum-ibid.\  {\bf 376}, 339 (2003)], hep-th/0204089.
\bibitem{ABS}
C.~Angelantonj, M.~Bianchi, G.~Pradisi, A.~Sagnotti and Y.~S.~Stanev,
Phys.\ Lett.\ B {\bf 385}, 96 (1996), hep-th/9606169.
\bibitem{CL}
M.~Cveti\v c and P.~Langacker,
Nucl.\ Phys.\ {\bf B586}, 287 (2000)
hep-th/0006049.

\bibitem{cim5}
D.~Cremades, L.~E.~Ib\'a\~nez and F.~Marchesano,
hep-ph/0212048.

\bibitem{cim6}
D.~Cremades, L.~E.~Ib\'a\~nez and F.~Marchesano,
JHEP {\bf 0307}, 038 (2003),
hep-th/0302105.


\bibitem{cpw}
M.~Cveti\v c, M.~Pl\"umacher and J.~Wang, JHEP {\bf 0004}, 004 (2000),
 hep-th/9911021.

\bibitem{WITT}
E.~Witten,
Phys.\ Lett.\ B {\bf 117} (1982) 324.

\bibitem{pdg}
S. Eidelman et al., Phys. Lett. {\bf B592}, 1 (2004).

\bibitem{Susskind:2004uv}
L.~Susskind,
arXiv:hep-th/0405189.

\bibitem{Douglas:2004qg}
M.~R.~Douglas,
hep-th/0405279.


\bibitem{Arkani-Hamed:2004fb}
N.~Arkani-Hamed and S.~Dimopoulos,
hep-th/0405159.
\bibitem{UR}
A.~M.~Uranga,
Nucl.\ Phys.\ B {\bf 598}, 225 (2001), hep-th/0011048.

\bibitem{MS}
F.~Marchesano and G.~Shiu,
arXiv:hep-th/0409132.





\end{thebibliography}
\end{document}